\documentclass{article}

\usepackage{PRIMEarxiv}

\usepackage[utf8]{inputenc} 
\usepackage[T1]{fontenc}    
\usepackage{hyperref}       
\usepackage{url}            
\usepackage{booktabs}       
\usepackage{amsfonts}       
\usepackage{nicefrac}       
\usepackage{microtype}      
\usepackage{lipsum}
\usepackage{fancyhdr}       
\usepackage{graphicx}       
\usepackage{defs}
\usepackage{amsmath}
\usepackage{amssymb}
\usepackage{subcaption}
\usepackage{tabularx}
\usepackage{booktabs}
\usepackage{listings}
\usepackage{xcolor}
\graphicspath{{Figures/}}     

\title{Data-driven Synthesis of Magnetic Resonance Spectroscopy Data using a Variational Autoencoder
\thanks{Pre-print: This manuscript has not been peer-reviewed.}}

\author{
Dennis M.J. van de Sande \\
Department of Biomedical Engineering \\
Eindhoven University of Technology \\
Eindhoven, The Netherlands \\
\texttt{d.m.j.v.d.sande@tue.nl} 
\And
Julian P. Merkofer \\
Department of Electrical Engineering \\
Eindhoven University of Technology \\
Eindhoven, The Netherlands \\
\And
Sina Amirrajab \\
Department of Biomedical Engineering \\
Eindhoven University of Technology \\
Eindhoven, The Netherlands \\
\And
Mitko Veta \\
Department of Biomedical Engineering \\
Eindhoven University of Technology \\
Eindhoven, The Netherlands \\
\And
Gerhard S. Drenthen \\
Department of Radiology \& Nuclear Medicine \\
Maastricht University Medical Center \\
Maastricht, The Netherlands \\
\And
Jacobus F.A. Jansen \\
Department of Electrical Engineering \\
Eindhoven University of Technology \\
Eindhoven, The Netherlands \\
\& Department of Radiology \& Nuclear Medicine \\
Maastricht University Medical Center \\
Maastricht, The Netherlands \\
\And
Marcel Breeuwer \\
Department of Biomedical Engineering \\
Eindhoven University of Technology \\
Eindhoven, The Netherlands \\
\& Department of Electrical Engineering \\
Eindhoven University of Technology \\
Eindhoven, The Netherlands \\
}

\date{}  

\begin{document}
\maketitle

\begin{abstract}
The development of deep learning methods for magnetic resonance spectroscopy (MRS) is often hindered by the limited availability of large, high-quality training datasets. While physics-based simulations are commonly used to mitigate this limitation, accurately modeling all in-vivo signal components remains challenging. In this work, we propose a data-driven framework for synthesizing in-vivo MRS data using a variational autoencoder (VAE) trained exclusively on measured single-voxel spectroscopy data. The model learns a low-dimensional latent representation of complex-valued spectra and enables the generation of new samples through latent-space sampling and interpolation strategies.
The generative performance of the proposed approach is evaluated using a comprehensive set of complementary analyses, including reconstruction quality, feature-level similarity using low-dimensional embeddings, application-based signal quality metrics, and metabolite quantification agreement. The results demonstrate that the VAE accurately reconstructs dominant spectral patterns and generates synthetic spectra that occupy the same feature space as in-vivo data. In an example application targeting GABA-edited spectroscopy, augmenting limited subsets of transients with synthetic spectra improves signal quality metrics such as signal-to-noise ratio, linewidth, and shape scores.
However, the results also reveal limitations of the generative approach, including under-representation of stochastic noise and reduced accuracy in absolute metabolite quantification, particularly for applications sensitive to concentration estimates. These findings highlight both the potential and the limitations of data-driven MRS synthesis. Beyond the proposed model, this study introduces a structured evaluation framework for generative MRS methods, emphasizing the importance of application-aware validation when synthetic data are used for downstream analysis.
\end{abstract}


\section{Introduction}\label{sec:introduction}

\Ac{ai} and \ac{dl} are increasingly being adopted in the field of MRS to perform a wide range of tasks, including spectral reconstruction, artifact removal, quantification, classification, and other applications \cite{chenReviewProspectDeep2020, vandesandeReviewMachineLearning2023}. However, the development of \ac{dl}-based methods typically requires large amounts of high-quality training data. In MRS, the availability of such data is limited due to the time-consuming nature of acquisitions, high acquisition costs, privacy concerns, and the lack of large open-source databases. This is partly because MRS is not routinely included in clinical imaging protocols.

To mitigate this data scarcity, researchers often rely on model-based MRS simulations to generate artificial training data. Several toolboxes have been developed for this purpose \cite{simpsonAdvancedProcessingSimulation2017, clarkeFSLMRSEndtoendSpectroscopy2021, soherVespaIntegratedApplications2023, lamasterMRSSimOpenSourceFramework2025, vandesandeDigitalPhantomMR2026}, most of which employ reverse-fitting approaches, whereby fitting parameters are provided to the signal model to synthesize spectra rather than being estimated from measured data. While metabolite signals can be accurately simulated using the density matrix formalism \cite{mulkernDensityMatrixCalculations1994}, other signal contributions, such as macromolecular background signals, residual water, lipids, and additional in-vivo nuisance components, remain challenging to model accurately. Incomplete or simplified modeling of these contributions may result in synthetic spectra that only partially capture the variability observed in practice, leading to domain shift and reduced generalization of \ac{dl} models trained on such data. Moreover, there is no broadly accepted consensus on the most appropriate signal models or parameter ranges for generating realistic MRS data, which has led to complex, simulation pipelines that are typically not publicly available, thereby hampering independent validation and reproducibility.

As an alternative to physics-based simulations, MRS data can also be generated using data-driven synthesis approaches. These methods leverage existing datasets to train generative models that learn the underlying data distribution, enabling the generation of new spectra without explicitly defining parameter ranges or complex forward models. Generative \ac{ai} techniques have been applied to various spectroscopy modalities \cite{flanaganExploringGenerativeArtificial2025}; nevertheless, only a limited number of studies have investigated their application to in-vivo MRS. One example is the work by Maruyama et al.~\cite{maruyamaGeneratingSyntheticMR2025}, who proposes a \ac{gan}-based framework for generating \ac{mrsi} data using complementary MRI and \ac{svs} information. Other studies employing \ac{gan} architectures include the work by Olliverre et al.~\cite{olliverreGeneratingMagneticResonance2018}, which focused on generating \ac{svs} data for different brain tumor grades and evaluated the synthetic spectra using a downstream classification task, as well as the study by Jang et al.~\cite{jangUnsupervisedAnomalyDetection2021}, which applied \ac{gan}-based models for anomaly detection. Notably, in the latter study, the model was trained exclusively on simulated data.

To date, generative \ac{ai} approaches for MRS have predominantly relied on \ac{gan}-based models, while alternative generative frameworks, such as \acp{vae} \cite{kingmaAutoEncodingVariationalBayes2014}, remain largely unexplored. In contrast to \acp{gan}, \acp{vae} provide a probabilistic generative framework with a structured latent space, which has been shown to facilitate the synthesis of realistic medical data and support data augmentation in various imaging domains, including segmentation and classification tasks \cite{kebailiDeepLearningApproaches2023, raisExploringVariationalAutoencoders2024}. Furthermore, there is currently no standardized or widely accepted framework for evaluating the realism and validity of synthetic MRS data, with existing studies employing heterogeneous and application-specific evaluation strategies. Therefore, the aim of this study is to develop a \ac{vae}-based framework for generating synthetic \ac{svs} MRS data using exclusively in-vivo data. In this work, the framework is developed and evaluated using in-vivo proton brain MRS data. In addition, the generated spectra are systematically evaluated using multiple complementary strategies, including metric-based comparisons, feature similarity analysis, quantification agreement, and application-based evaluations.

\section{Materials \& Methods}\label{sec:materials&methods}
\subsection{Data}
The dataset used in this study is \ac{svs} MRS data acquired from a subcohort of The Maastricht Study dataset\cite{schramMaastrichtStudyExtensive2014}, and is previously used in van Bussel et al. \cite{vanbusselIncreasedGABAConcentrations2016}. The Maastricht Study was approved by the Medical Ethics Committee of the Maastricht University Medical Center+ (Maastricht, the Netherlands), and all participants provided written informed consent. The dataset used in this study comprises 102 subjects in total. The mean age across all participants is $62.2 \pm 8.0$ years, with 59 men and 43 women. The cohort is divided into three groups: healthy controls (Group C, $N=41$, age $59.3 \pm 9.0$ years, 15 men/26 women), patients with type 2 diabetes (Group D, $N=44$, age $64.8 \pm 6.0$ years, 33 men/11 women), and patients with metabolic syndrome (Group M, $N=17$, age $62.6 \pm 8.0$ years, 11 men/6 women). More details on the patient population can be found in van Bussel et al. \cite{vanbusselIncreasedGABAConcentrations2016}.

MRS data is acquired using a 3T scanner (Achieva TX, Philips Healthcare, Best, The Netherlands) and a \ac{mega-press} sequence (TR/TE = 2000/68ms, editing pulses at 1.9 (ON), and 7.46 ppm (OFF), MOIST water suppression, 10:40 minutes acquisition time). The measured voxel is located in the occipital lobe and has a size of $3\times3\times3$ cm$^3$. The \ac{nsa} is 320, interleaved in 40 ON/OFF blocks with 8 \ac{nsa} each. Since individual transients are not accessible in the given data format (.SPAR/.SDAT), one block of 8 \ac{nsa} is defined as a single transient for the rest of this study. All spectra contain 2048 spectral points with a bandwidth of 2000 Hz. 

All files are converted to the NIfTI-MRS format using spec2nii \cite{clarkeNIfTIMRSStandardData2022}. \Acp{fid} are preprocessed by applying frequency and phase correction to align the ON and OFF transients using FSL-MRS \cite{clarkeFSLMRSEndtoendSpectroscopy2021}. This phase and frequency correction is applied within the frequency search range of 1.9 to 4.2 ppm. Next, the data are Fourier transformed and normalized by dividing the magnitude spectrum by its maximum.  The ON and OFF and the real and imaginary parts of the spectra are split into separate dimensions for further use in \ac{dl} models, making the dimensions for one subject: (20, 2, 2, 2048) where the dimensions represent the number of transients, ON/OFF channel, real/imaginary channel, and number of spectral points, respectively. The dataset was split at the subject level into training ($70\%$), validation ($15\%$), and test ($15\%$) sets using stratified sampling based on diagnostic group (healthy control, type 2 diabetes, metabolic syndrome). All transients from a given subject were assigned exclusively to a single split. No explicit quality-based exclusion is performed, as minor imperfections such as suboptimal water suppression or mild instabilities are intentionally retained to reflect realistic in-vivo conditions. Visual inspection confirmed that no spectra contained severe artifacts that could compromise analyses.


\subsection{\Ac{vae} Model}\label{subsec:model}
A \ac{vae} is used to model and generate complex-valued MRS spectra by learning a low-dimensional latent representation of the spectral data. The \ac{vae} is an unsupervised \ac{dl} model consisting of two \acp{nn}: an encoder $Q(Z|X)$ and a decoder $P(X|Z)$. The encoder maps an input spectrum $X$ to a latent representation $Z$, while the decoder reconstructs the spectrum as $\hat{X}$ from samples drawn from the latent space. MRS spectra are inherently complex-valued and consist of real and imaginary components. Both components are treated as separate input channels and are used simultaneously in all loss computations. Each spectrum therefore has the form 
\[
X = (X^{\mathrm{Re}}, X^{\mathrm{Im}}),
\] 
with corresponding decoder outputs 
\(\hat{X} = (\hat{X}^{\mathrm{Re}}, \hat{X}^{\mathrm{Im}})\). Each component contains \(N\) spectral points, such that
\[
X^{\mathrm{Re}} = (x_1^{\mathrm{Re}}, \ldots, x_N^{\mathrm{Re}}), 
\quad 
X^{\mathrm{Im}} = (x_1^{\mathrm{Im}}, \ldots, x_N^{\mathrm{Im}}),
\]
and analogously for \(\hat{X}^{\mathrm{Re}}\) and \(\hat{X}^{\mathrm{Im}}\). All subsequent loss computations are applied to both channels simultaneously.

\subsubsection{Loss Function}\label{subsubsec:loss_function}
The model is trained by minimizing a total loss that combines a reconstruction term and a latent regularization term: 

\begin{equation}
    \mathcal{L}_{total} = \mathcal{L}_{recon}(X, \hat{X}) + \beta_{KL}\mathcal{L}_{reg}(Q(Z|X), P(Z)) 
\end{equation}

where $\mathcal{L}_{recon}$ measures the reconstruction fidelity, $\mathcal{L}_{reg}$ penalizes deviations of the latent distribution from a prior $P(Z)$, and $\beta_{KL}$ controls the relative contribution of the regularization term. Latent regularization is achieved using the standard \ac{kl} divergence between the approximate posterior $Q(Z|X)$, parameterized by $(\mu,\text{log } \sigma^2)$, and an isotropic Gaussian prior $P(Z)=\mathcal{N}(0,I)$. Mathematically, this can be written as

\begin{equation}
    \mathcal{L}_{reg}(Q(Z|X), P(Z))=-\frac{1}{2}\sum_{j=1}^d(1+\text{log }\sigma_j^2-\mu_j^2-\sigma_j^2),
\end{equation}

with $d$ the dimensionality of the latent space. This term encourages a smooth, well-structured latent space suitable for generative sampling.

Reconstruction accuracy is enforced using a weighted point-wise loss designed to capture metabolite peaks, \ac{mm} and baseline contributions, as well as residual lipid and water signals. This loss combines \ac{mse} and L1 terms, with higher weights assigned to the primary spectral region of interest and lower weights elsewhere. With the spectral axis in ppm defined as $(p_1,\dots,p_N)$ for a spectrum with $N$ points, a per-point weighting factor is defined as:

\begin{equation}
  w_{i} =
    \begin{cases}
      w_{high}, & p_{min} \leq p_i \leq p_{max} \\
      w_{low}, & \text{otherwise},\\
    \end{cases}       
\end{equation}

where $(p_{min}, p_{max})$ defines the weighted spectral region (set to $0$–$7$ ppm in this work), $w_{high}$ is the weight used within this range, and $w_{low}$ is the weight applied outside this region. Using these weights, the point-wise loss is defined as:

\begin{equation}
    \mathcal{L}_{points}(X,\hat{X})=\alpha_{L2}\frac{1}{N}\sum^N_{i=1}w_i(\hat{x}_i-x_i)^2+\alpha_{L1}\frac{1}{N}\sum^N_{i=1}w_i|\hat{x}_i-x_i| 
\end{equation}

where $\alpha_{L2}$ and $\alpha_{L1}$ control the relative contribution of the \ac{mse} and L1 terms, respectively. The \ac{mse} term promotes accurate reconstruction of dominant peaks but can overemphasize large-amplitude deviations, whereas the L1 term provides a more uniform penalty across residuals. Their combination balances peak fidelity with robust reconstruction of lower-amplitude spectral structure. While this point-wise loss emphasizes accurate modeling of the signals of interest, it does not constrain the structure of the residual signal ($r=\hat{X}-X$). Structured residuals may indicate systematic reconstruction errors, such as baseline distortions or unresolved spectral features.
To mitigate this, a \ac{fft}-based residual loss is introduced. Although the residual $r$ is already represented in the frequency domain, we compute the \ac{fft}, denoted by $\mathcal{F}(\cdot)$, of the complex residual and evaluate the variance of its magnitude, $\sigma^2_{|\mathcal{F}(r)|}$. This metric penalizes coherent or periodic structure across spectral indices rather than random noise, thereby encouraging the residual to be noise-like instead of exhibiting low- or high-frequency patterns that may indicate systematic reconstruction errors.
Therefore, the final reconstruction loss is defined as

\begin{equation}
    \mathcal{L}_{recon}(X,\hat{X})=\mathcal{L}_{points}(X,\hat{X})+\alpha_{\mathcal{F}}\sigma^2_{\left|\mathcal{F}(r)\right|},
\end{equation}

where $\alpha_{\mathcal{F}}$ controls the relative contribution of the \ac{fft}-based residual loss.

\subsubsection{Model Architecture}\label{subsubsec:model_architecture}
The model architecture consists of symmetric encoder and decoder networks built from fully connected layers. Each input spectrum is represented as a complex-valued signal with length $N=2048$ points, stored as two channels corresponding to the real and imaginary parts. Before entering the encoder, one transient of either the ON or OFF channel is selected. 

\begin{figure}
    \centering
    \includegraphics[width=\linewidth]{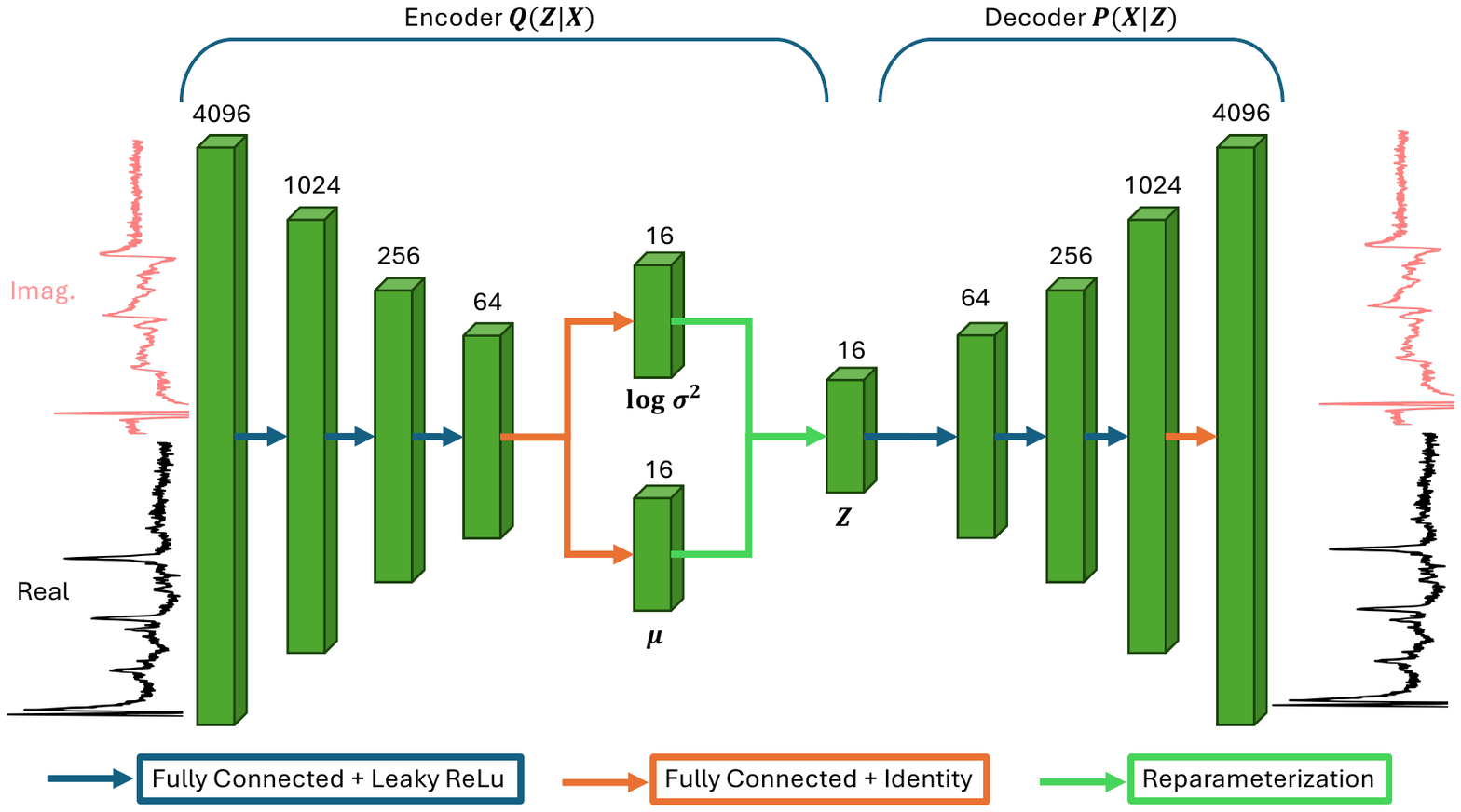}
    \caption{Overview of the used model architecture. Real and imaginary parts of the spectrum are concatenated and used as input for a fully connected \ac{nn}}
    \label{fig:model_architecture}
\end{figure}

During training, data augmentation is applied in the form of random frequency and phase shifts to enhance training data variability and improve robustness to small spectral variations. Phase shifts are sampled from a normal distribution with a standard deviation of $\pi/4$ radians, and frequency shifts are sampled from a normal distribution with a standard deviation of 10 Hz, applied along the spectral axis.
The two-channeled spectra are then flattened into a single vector of dimension $2N=4096$. Three fully connected layers in the encoder map this input vector into a low-dimensional latent representation. The decoder mirrors the encoder, expanding the latent vector back into the spectral domain. An overview of the model architecture is shown in Figure \ref{fig:model_architecture}.

The model is trained using the Adam optimizer with a learning rate scheduler that reduces the learning rate when the validation loss is not improving for 10 epochs. Early stopping is applied if the validation loss does not improve for 30 epochs. Training is performed on a Linux cluster node with 64 CPU cores and one NVIDIA H100 GPU, using Python 3.11.14. The total training time for the final model is approximately 6.5 minutes. The complete configuration, including data preprocessing, model architecture, training procedure, and loss parameters, is provided in the accompanying public GitHub repository (see Code Availability statement).

\subsection{Data Generation}\label{subsec:data_generation}
\begin{figure}
    \centering
    \includegraphics[width=\linewidth]{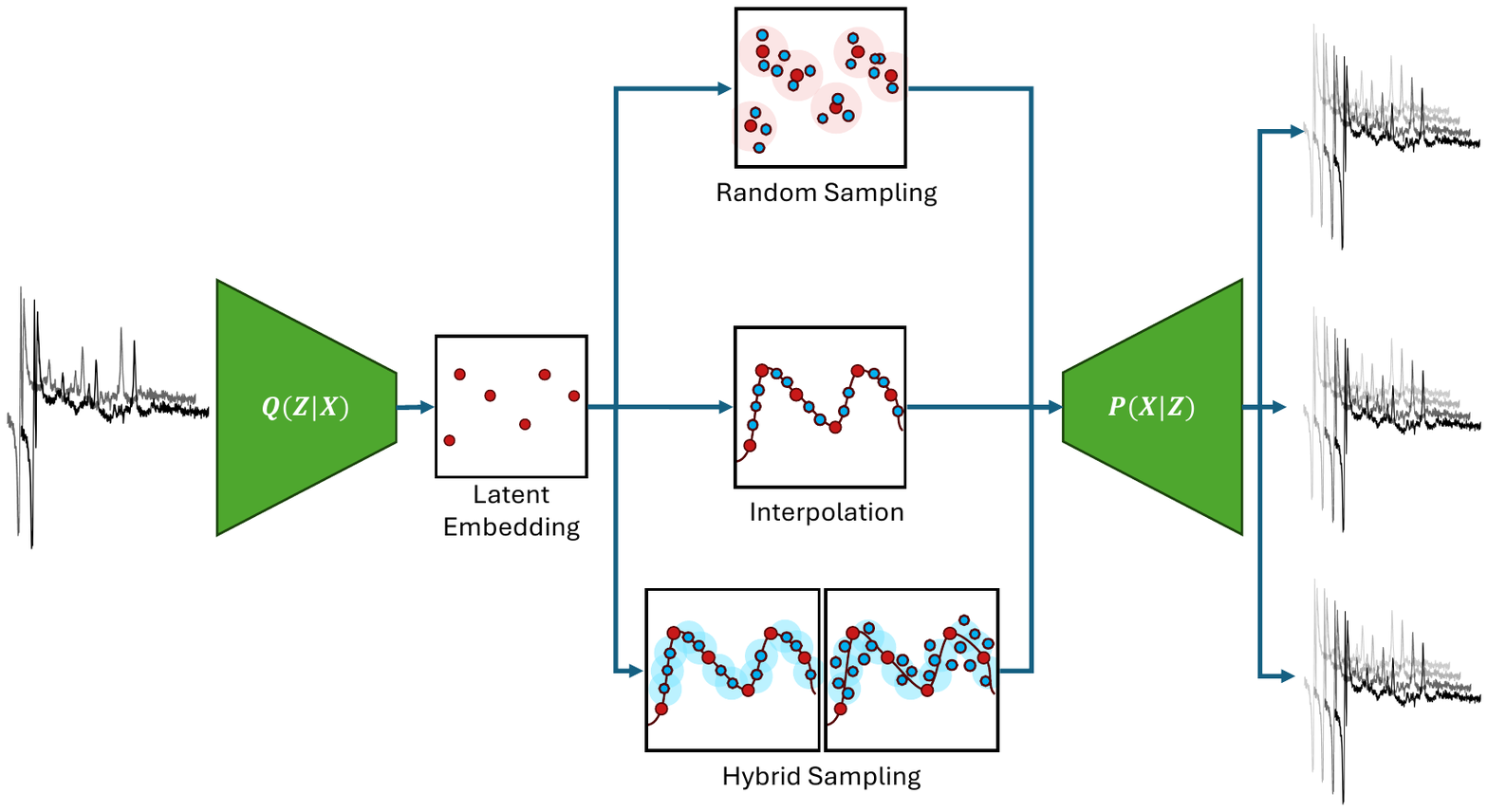}
    \caption{Overview of the data generation pipeline. Real, in vivo spectra are first encoded into latent space. New samples are then generated using one of three methods: random sampling, interpolation, or hybrid sampling. Finally, all generated samples are decoded back into spectra.}
    \label{fig:data_gen_pipeline}
\end{figure}

Once the \ac{vae} model has been trained, new data can be generated. Three different methods are used for the generation of new data, which involves the manipulation of encoded spectra within the latent space. A comprehensive overview of the entire process is presented in Figure \ref{fig:data_gen_pipeline}.

The first data generation technique involves random sampling in the learned latent space. In this method, in-vivo spectra are first encoded into latent vectors using the encoder \ac{nn}, and then random Gaussian noise is added to these latent vectors to produce variations. The perturbed latent vectors are subsequently decoded to generate new spectra. In this study, the Gaussian noise has a mean of 0 and a standard deviation of 0.1. This approach allows controlled variation around in-vivo spectra while remaining rooted in their underlying distribution, as visualized in Figure \ref{fig:data_gen_pipeline}.

Another technique is interpolation. Encoded latent vectors are obtained by encoding in-vivo spectra with the encoder \ac{nn}. New latent vectors are then generated by linear interpolation along each dimension between pairs or groups of encoded vectors. Decoding these interpolated vectors produces spectra that show gradual transitions in the features present in the original data. Linear interpolation is used in this study for all latent vector dimensions.

The final technique is a hybrid method that combines interpolation with random perturbations. Interpolated latent vectors are perturbed with Gaussian noise (using the same parameters as in the random sampling procedure) before decoding, thereby increasing the diversity of the generated spectra while still remaining anchored in the distribution of in-vivo spectra. All parameters for random perturbation, interpolation, and hybrid sampling, including Gaussian noise magnitudes and interpolation settings, are also provided in the accompanying public GitHub repository (see Code Availability statement).

\subsection{Evaluation}\label{subsec:evaluation}
\subsubsection{Generative Performance}
The generative capabilities of the developed \ac{vae} model are evaluated by assessing both the reconstruction quality of in-vivo spectra and the similarity between in-vivo and synthetic spectra. Reconstruction performance is examined through visual inspection and by comparing the \ac{snr} and linewidth metrics between the in-vivo test spectra and their corresponding reconstructions. The \ac{snr} is defined as the ratio between the maximum amplitude of the \ac{cr} resonance at approximately 3.0 ppm and twice the standard deviation of the spectral noise. Noise is estimated from a metabolite-free region between 9.8 and 10.8 ppm after removal of slow baseline trends using a second-order polynomial fit. The linewidth is defined as the \ac{fwhm} of the same \ac{cr} resonance.

Synthetic samples are subsequently generated from the test set using the three data-generation strategies described in Section \ref{subsec:data_generation}. To assess feature-level similarity between in-vivo and synthetic spectra, \ac{umap}\cite{mcinnesUMAPUniformManifold2020} is employed. All ON and OFF transients from the in-vivo test set are first used to train the \ac{umap} model and obtain a two-dimensional embedding of the complex spectra. Next, for each generation strategy, the synthetic spectra are projected into the same embedding space using the trained \ac{umap} model. Comparison of the resulting overlap and distribution of data points provides a measure of how well the synthetic spectra capture the feature characteristics of the in-vivo data.

\subsubsection{Application-based evaluation}\label{subsubsec:application_based_evaluation}
\begin{figure}
    \centering
    \includegraphics[width=\linewidth]{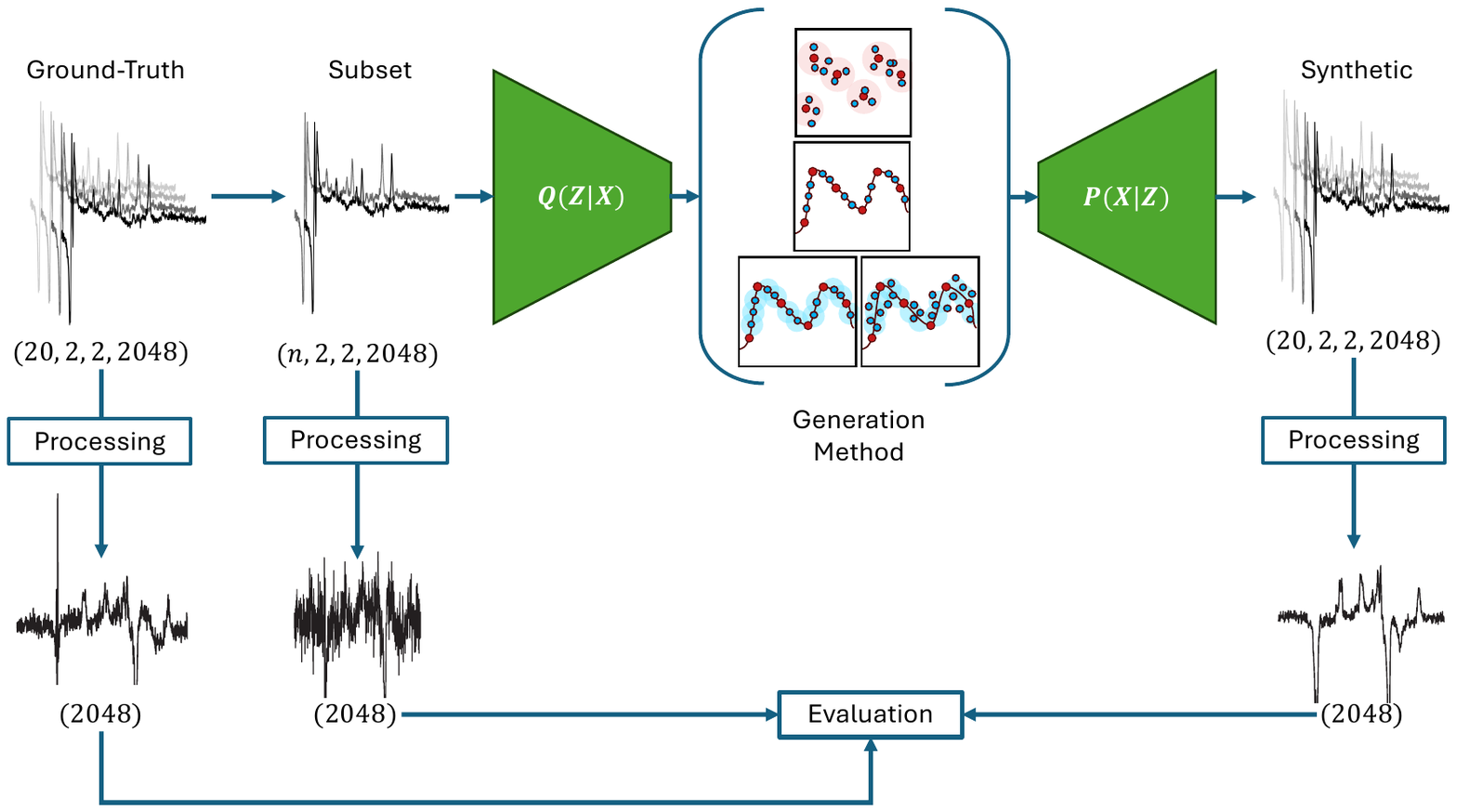}
    \caption{Overview of the application-based evaluation method. \Ac{gt} spectra from a single subject are used to select a subset of $n$ transients. These transients are encoded using the encoder \ac{nn} and used for the generation of synthetic samples. The decoder \ac{nn} generates synthetic transients to end up with the same number of transients as in the \ac{gt} spectra. For the evaluation all three datasets are processed and fitted using Osprey.}
    \label{fig:application_pipeline}
\end{figure}

In addition to the previously described evaluations, the practical usability of the generative model is demonstrated through a proof-of-concept application. The application, shown in Figure \ref{fig:application_pipeline}, focuses on spectral reconstruction of \ac{gaba}-edited spectra. The aim is to reduce scan time while maintaining spectral quality and quantitative accuracy by artificially increasing the number of transients using generative \ac{dl}.

The procedure begins with all transients from a single test subject, from which a subset of $n$ transients is selected. These subsets are obtained using a sliding-window technique that selects consecutive, non-overlapping transients to preserve the temporal relationships among them. Each subset is then encoded using the encoder \ac{nn}, and the resulting latent vectors are used to generate synthetic samples in latent space. These synthetic samples are subsequently decoded using the decoder \ac{nn} to create a synthetic dataset. In this study, $n$ is set to $2$. For each subset, 36 synthetic transients are generated ($2\times18$, ON and OFF), and the corresponding original in-vivo transients are added, resulting in a total of $40$ transients per subject that matches the ground-truth dataset.

After generation, the ground truth, subset, and synthetic spectra are processed in Osprey \cite{oeltzschnerOspreyOpensourceProcessing2020} to obtain \ac{gaba}-edited difference spectra. The efficacy of the \ac{vae} model in this context is assessed by comparing the \ac{snr}, linewidth, \ac{mse}, and the \ac{gaba} and \ac{glx} shape scores of all difference spectra. These selected metrics follow those used in the ISBI Edited MRS Reconstruction Challenge, as described by Berto et al. \cite{bertoResults2023ISBI2024}, and all focus on the \ac{gaba} and \ac{glx} peaks. Statistical significant differences between the calculated metrics are determined using a pairwise Wilcoxon Signed-Rank test. Finally, Osprey is used to fit the spectra and to compare \ac{gaba} and \ac{glx} concentrations from the difference spectra, as well as \ac{tnaa} and \ac{tcr} concentrations from the OFF spectra. Since no water reference is available for the synthetic data, the amplitudes from the fitting model will be used for comparison. After generation, the ground truth, subset, and synthetic spectra are processed in Osprey \cite{oeltzschnerOspreyOpensourceProcessing2020} to obtain \ac{gaba}-edited difference spectra. The efficacy of the \ac{vae} model in this context is assessed by comparing the \ac{snr}, linewidth, \ac{mse}, and the \ac{gaba} and \ac{glx} shape scores of all difference spectra. These selected metrics follow those used in the ISBI Edited MRS Reconstruction Challenge, as described by Berto et al. \cite{bertoResults2023ISBI2024}, and all focus on the \ac{gaba} and \ac{glx} peaks. Statistical significant differences between the calculated metrics are determined using a pairwise Wilcoxon Signed-Rank test. Finally, Osprey is used to fit the spectra and to compare \ac{gaba} and \ac{glx} concentrations from the difference spectra, as well as \ac{tnaa} and \ac{tcr} concentrations from the OFF spectra. Since no water reference is available for the synthetic data, the amplitudes from the fitting model will be used for comparison. Any residual water signal is removed by Osprey prior to metabolite fitting, ensuring that remaining water components do not influence quantification. All Osprey job-file settings are provided in appendix \ref{app:osprey_job}.

\section{Results}\label{sec:results}
All results are obtained using the final model configuration described in the Section \ref{subsec:model}. The complete set of hyperparameters and implementation details is publicly available in the accompanying GitHub repository.
\subsection{Generative Performance}\label{subsec:res_generative_performance}
\begin{figure}
    \centering
    \includegraphics[width=\linewidth]{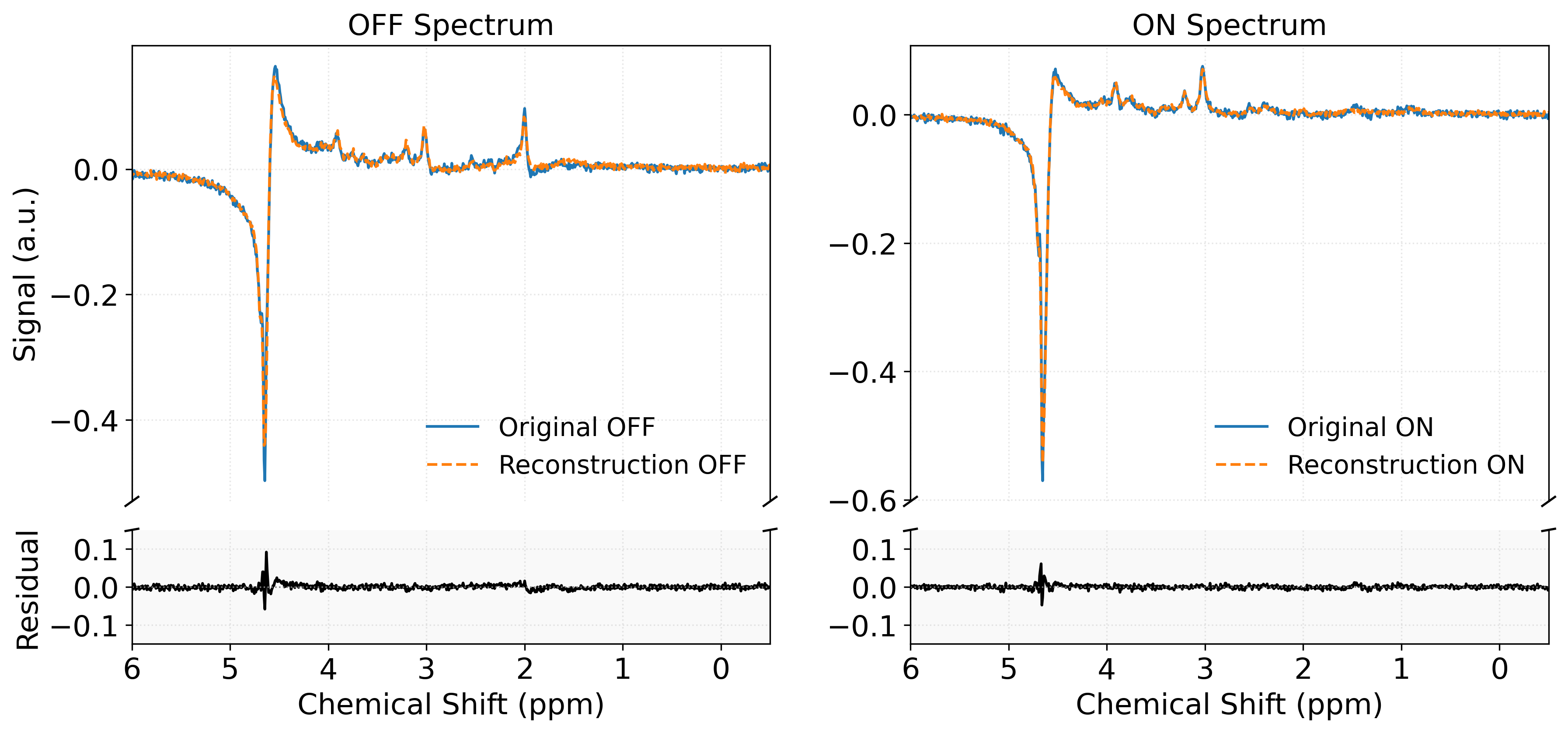}

    \vspace{0.5cm} 

    \includegraphics[width=\linewidth]{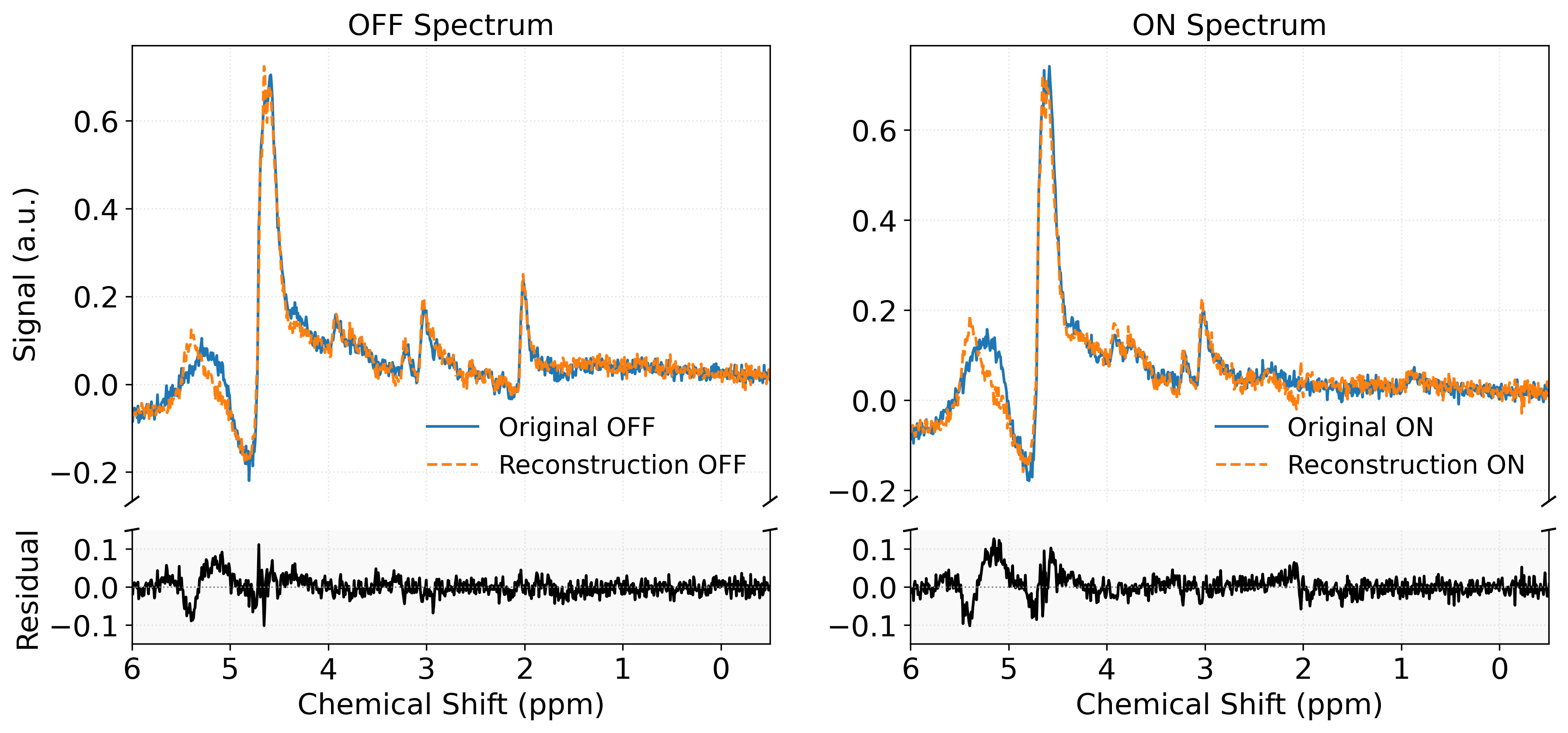}
    \caption{Representative in-vivo spectra (blue) and corresponding reconstructions (orange) with residuals (black) for two subjects. Top panels show OFF and ON spectra with small, flat residuals, indicating accurate reconstruction. Bottom panels show another subject with slightly larger residuals, primarily around the water signal, while the overall spectral shapes are preserved.}
    \label{fig:reconstructions}
\end{figure}

Figure \ref{fig:reconstructions} illustrates representative examples of in-vivo spectra together with their corresponding reconstructions and residuals. Spectra from two different subjects are shown, each including both OFF and ON transients. These examples are selected as representative of the overall reconstruction performance. Visual inspection of the top two spectra (OFF and ON) shows small and relatively flat residuals, indicating good reconstruction quality. Only minor deviations are observed around the residual water signal. The bottom two spectra exhibit slightly larger residual amplitudes, again predominantly near the water resonance. Importantly, in all cases the residuals do not show structured spectral patterns, suggesting that the reconstruction error is primarily driven by stochastic noise rather than systematic modeling errors.

Quantitative analysis confirms this observation. The reconstruction \ac{snr} is consistently higher than the in-vivo SNR, with values of $9.98 \pm 1.18$ (OFF) and $11.06 \pm 1.35$ (ON) for the in-vivo spectra, increasing to $17.53 \pm 8.36$ (OFF) and $19.05 \pm 9.01$ (ON) for the reconstructed spectra. This \ac{snr} increase is consistent with the visual appearance in Figure \ref{fig:reconstructions} and reflects that the \ac{vae} model suppresses high-frequency noise components rather than reproducing individual noise realizations present in the in-vivo data.

In contrast, the Cr linewidths are well preserved by the reconstruction process. The in-vivo linewidths are $0.057 \pm 0.017$ ppm (OFF) and $0.062 \pm 0.018$ ppm (ON), compared to $0.057 \pm 0.019$ ppm (OFF) and $0.063 \pm 0.017$ ppm (ON) for the reconstructed spectra. This close agreement indicates that the spectral lineshapes are accurately reconstructed despite the reduction in noise.
 
\begin{figure*}
    \centering
    \includegraphics[width=\linewidth]{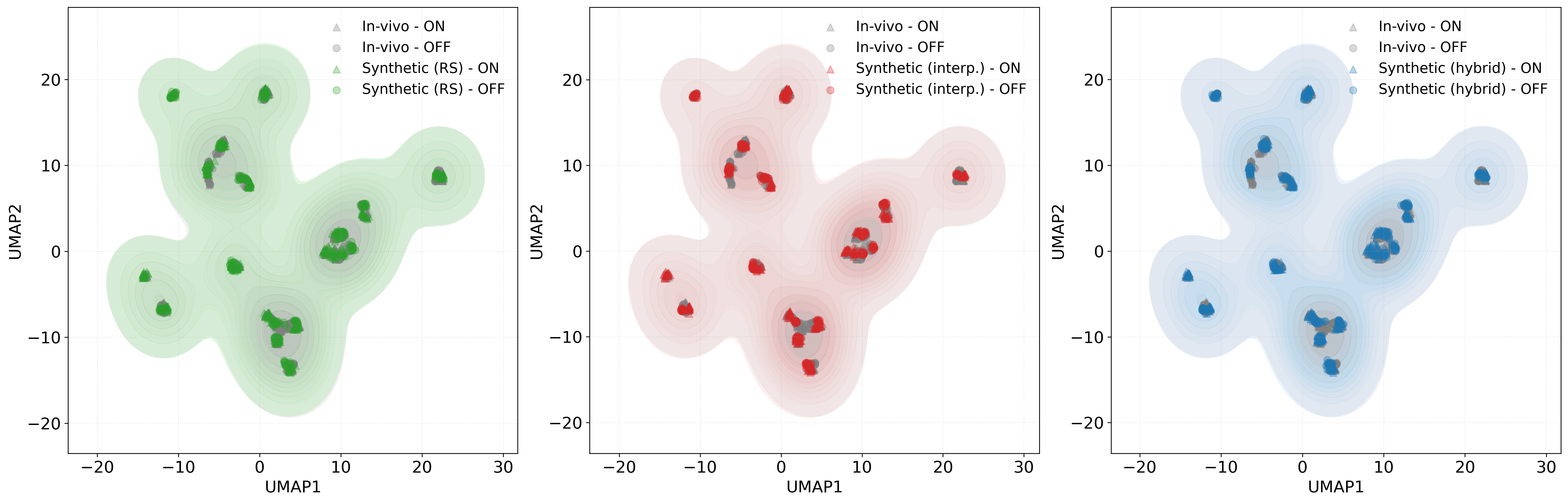}
    \caption{\ac{umap} visualizations comparing \ac{gt} in-vivo spectra with synthetic spectra generated using three different data generation methods: random sampling (RS) (green), interpolation (red), and hybrid sampling (blue). The \ac{umap} embedding is learned using the \ac{gt} data, after which synthetic spectra generated from a single subset of transients are projected into the same low-dimensional space. Each point represents an individual spectrum, with markers indicating ON and OFF acquisitions. The shaded regions represent kernel density estimates of the embedded spectral distributions for both \ac{gt} and synthetic data, illustrating the overall spatial extent and degree of overlap between datasets. Clusters corresponding to individual subjects are visible across all panels.}
    \label{fig:umap}
\end{figure*}

Figure \ref{fig:umap} shows the \ac{umap} visualizations comparing the three data generation methods with the \ac{gt} in-vivo dataset. To limit visual clutter, only synthetic spectra generated from a single subset of transients are shown. Across all panels, distinct clusters corresponding to individual subjects are observed. The synthetic spectra largely overlap with the in-vivo data, indicating a high degree of feature similarity between synthetic and real spectra in the embedded space. No pronounced differences between the generation methods are observed. Methods that incorporate random sampling of latent vectors show, at most, a slightly increased spread in the \ac{umap} space.

\subsection{Application-based evaluation}
\begin{figure*}
    \centering
    \begin{subfigure}{0.3\linewidth}
        \centering
        \includegraphics[width=\linewidth]{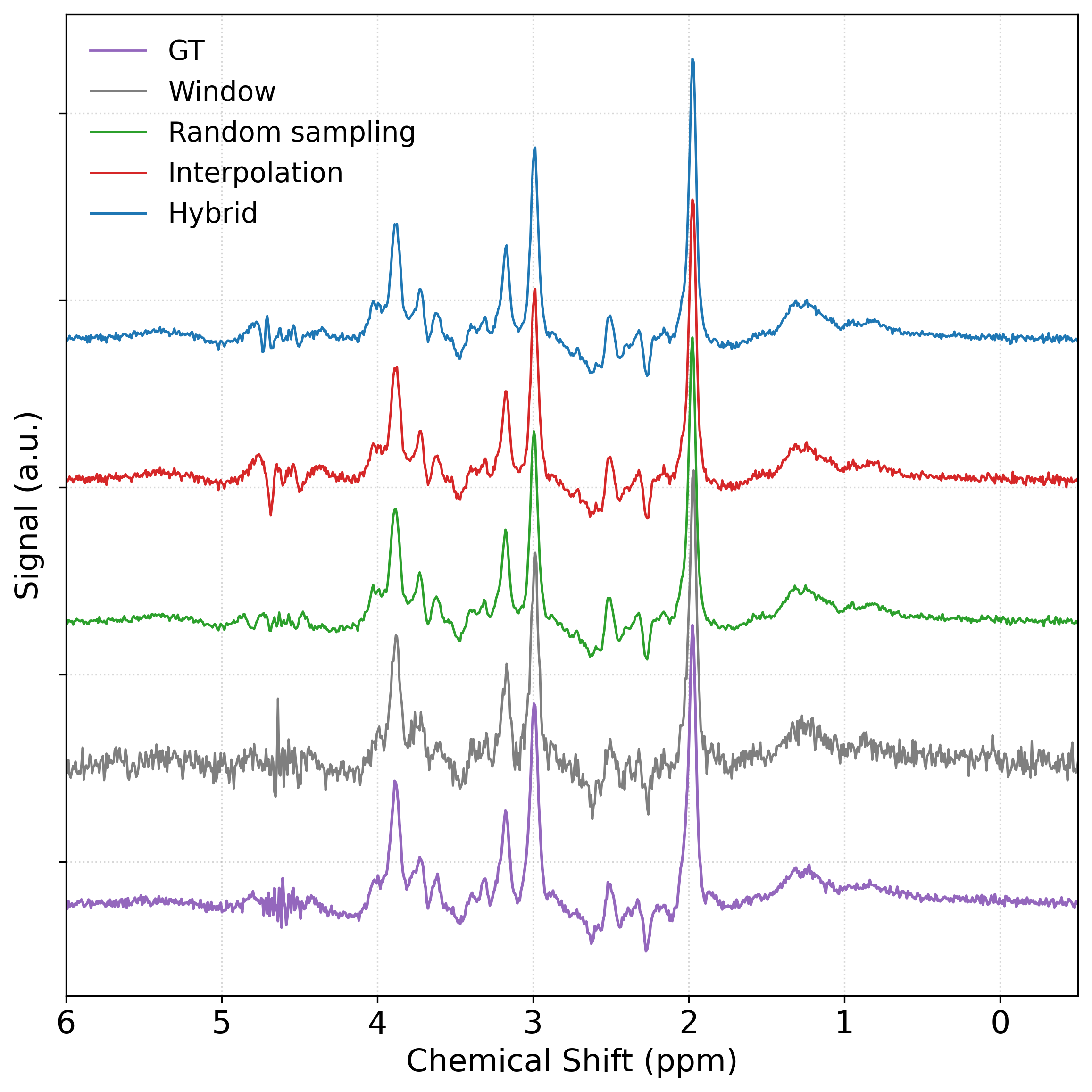}
        \caption{OFF spectra for all methods, subject C008.}
    \end{subfigure}
    \begin{subfigure}{0.3\linewidth}
        \centering
        \includegraphics[width=\linewidth]{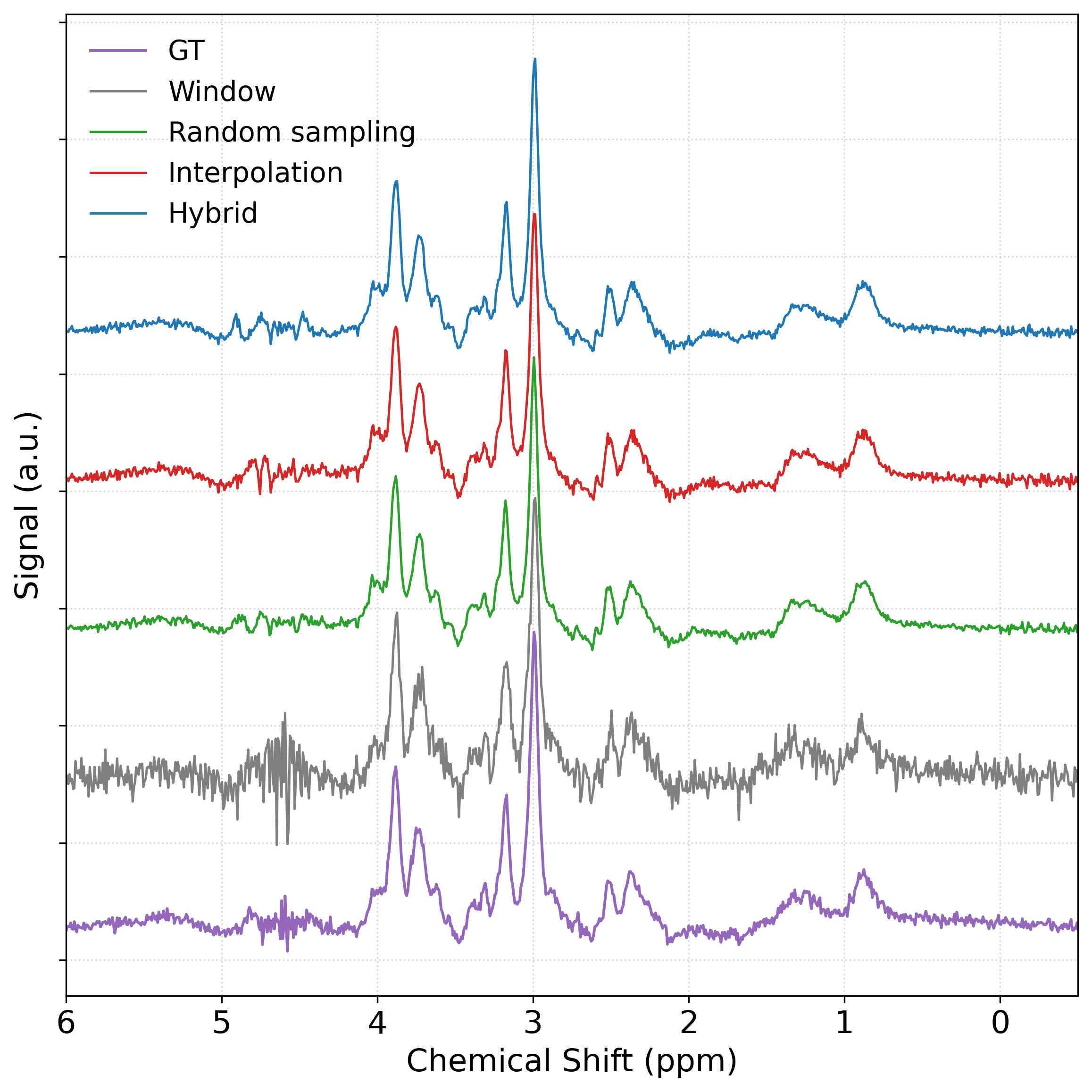}
        \caption{ON spectra for all methods, subject C008.}
    \end{subfigure}
    \begin{subfigure}{0.3\linewidth}
        \centering
        \includegraphics[width=\linewidth]{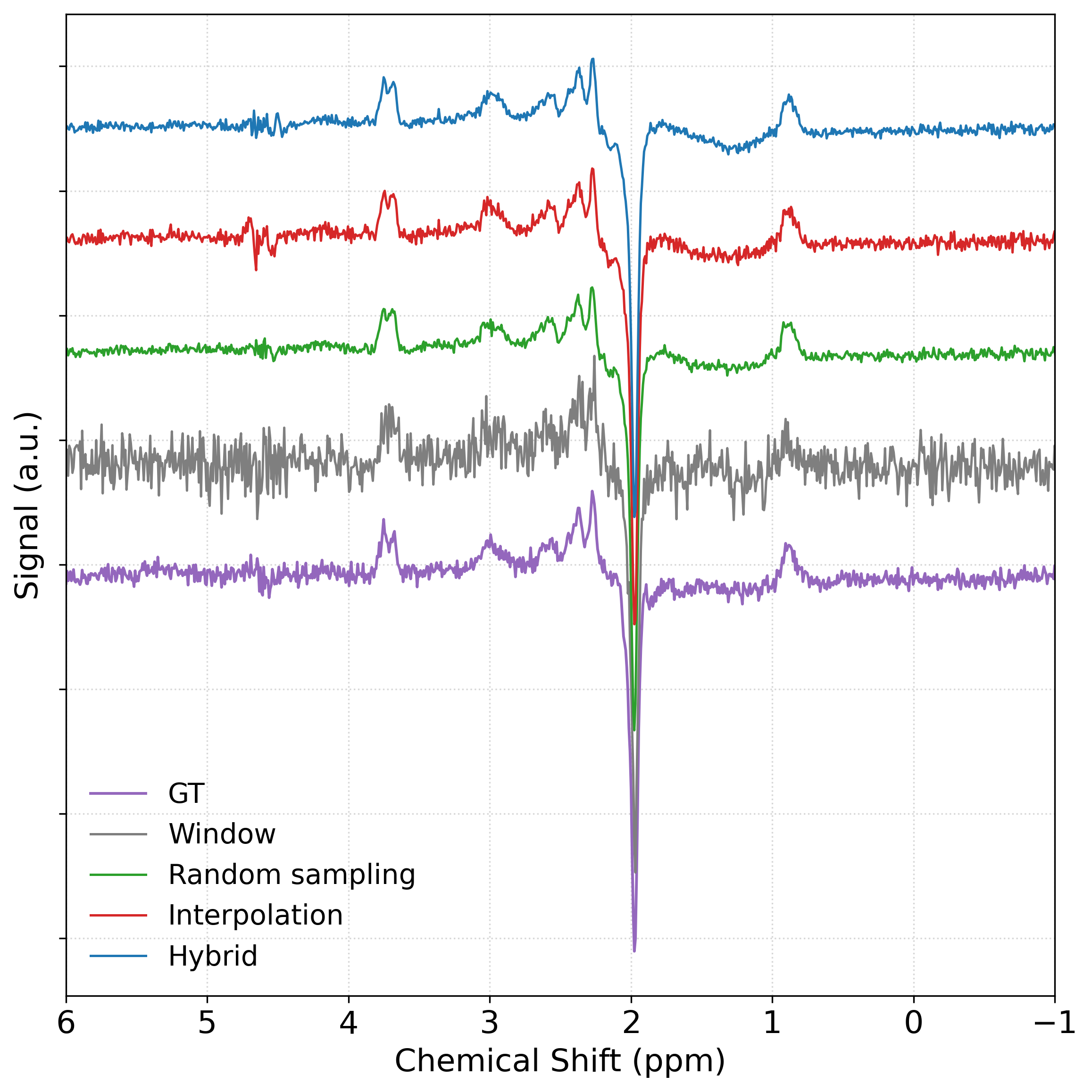}
        \caption{Difference (OFF-ON) spectra for all methods, subject C008.}
    \end{subfigure}

    \vspace{0.5em} 

    \begin{subfigure}{0.3\linewidth}
        \centering
        \includegraphics[width=\linewidth]{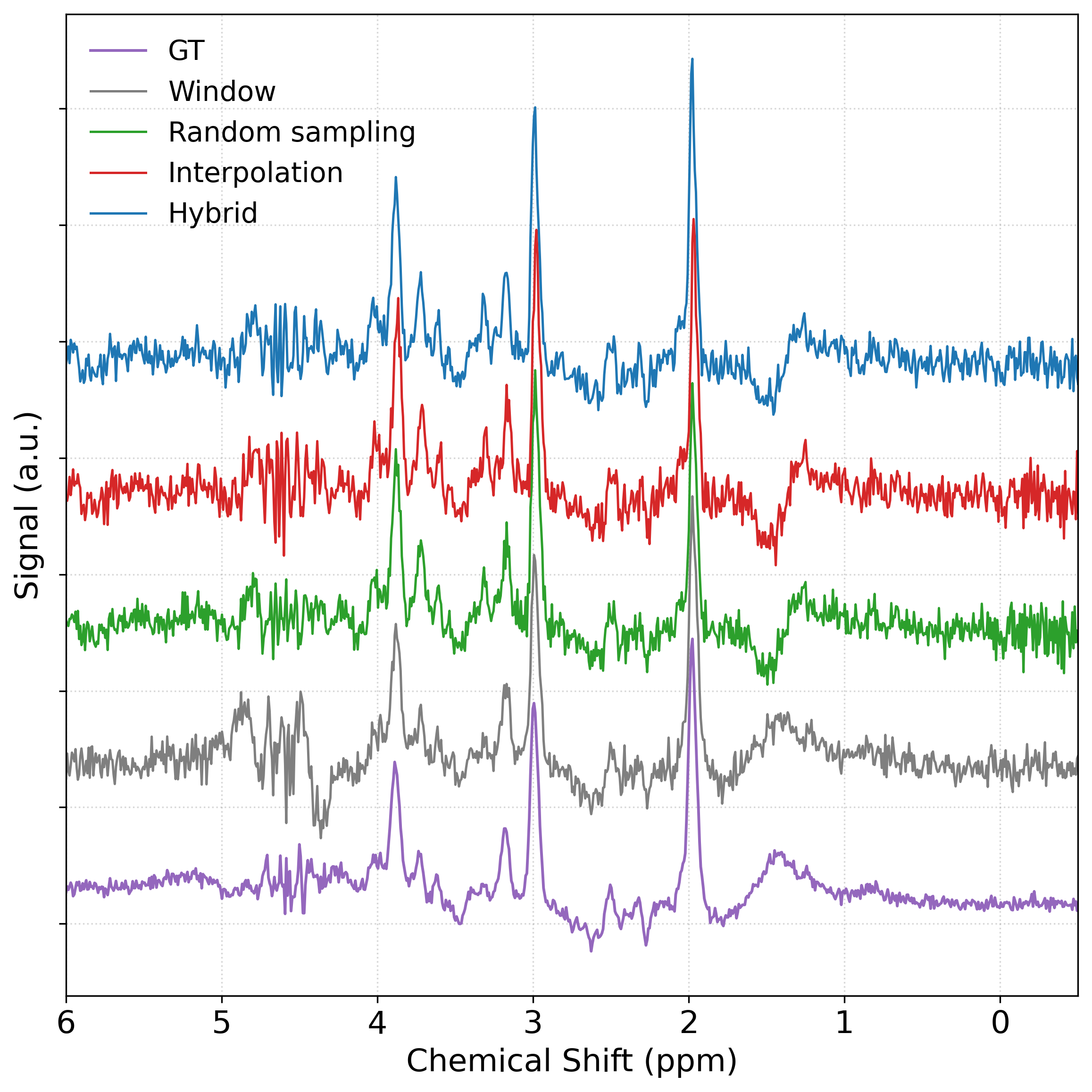}
        \caption{OFF spectra for all methods, subject C039.}
    \end{subfigure}
    \begin{subfigure}{0.3\linewidth}
        \centering
        \includegraphics[width=\linewidth]{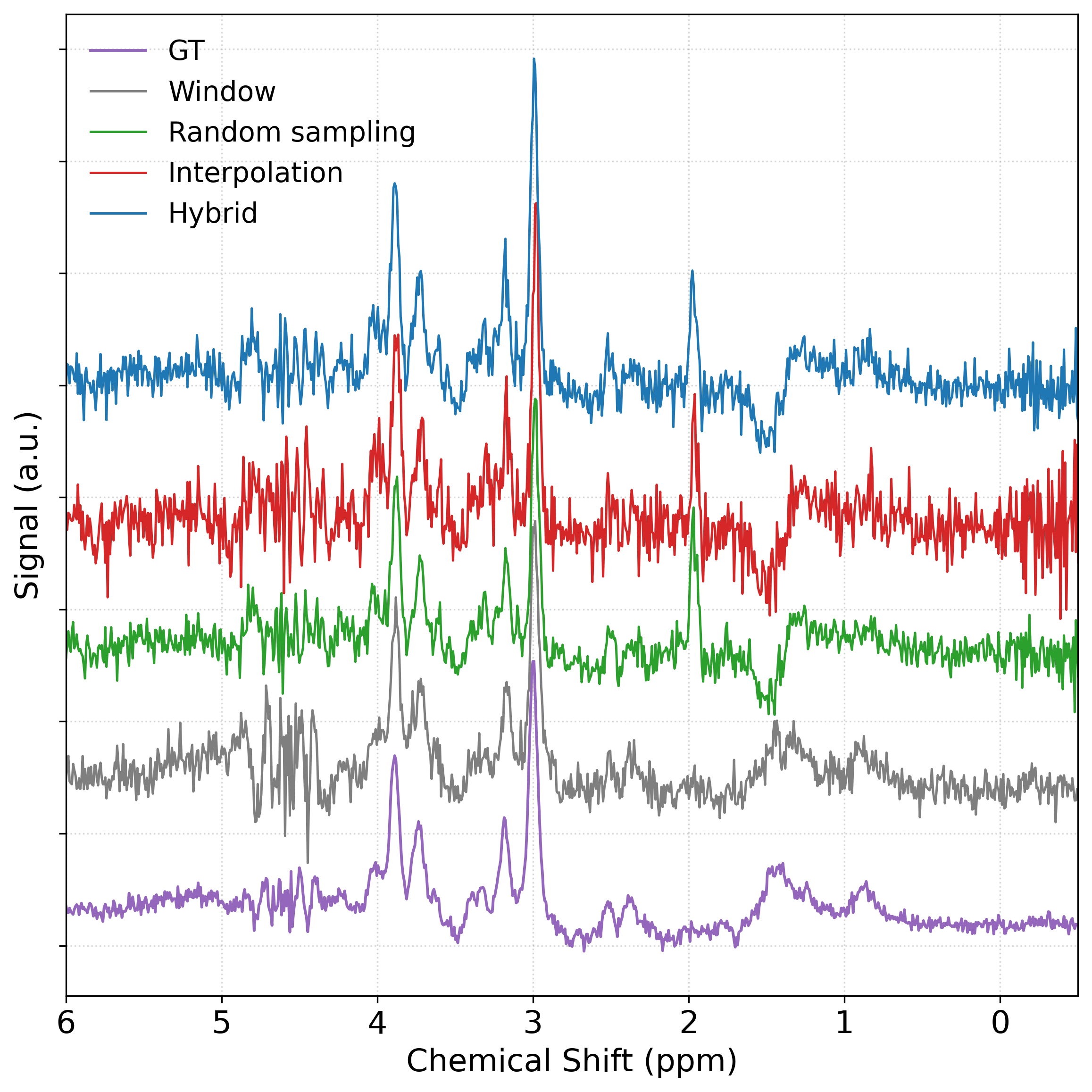}
        \caption{ON spectra for all methods, subject C039.}
    \end{subfigure}
    \begin{subfigure}{0.3\linewidth}
        \centering
        \includegraphics[width=\linewidth]{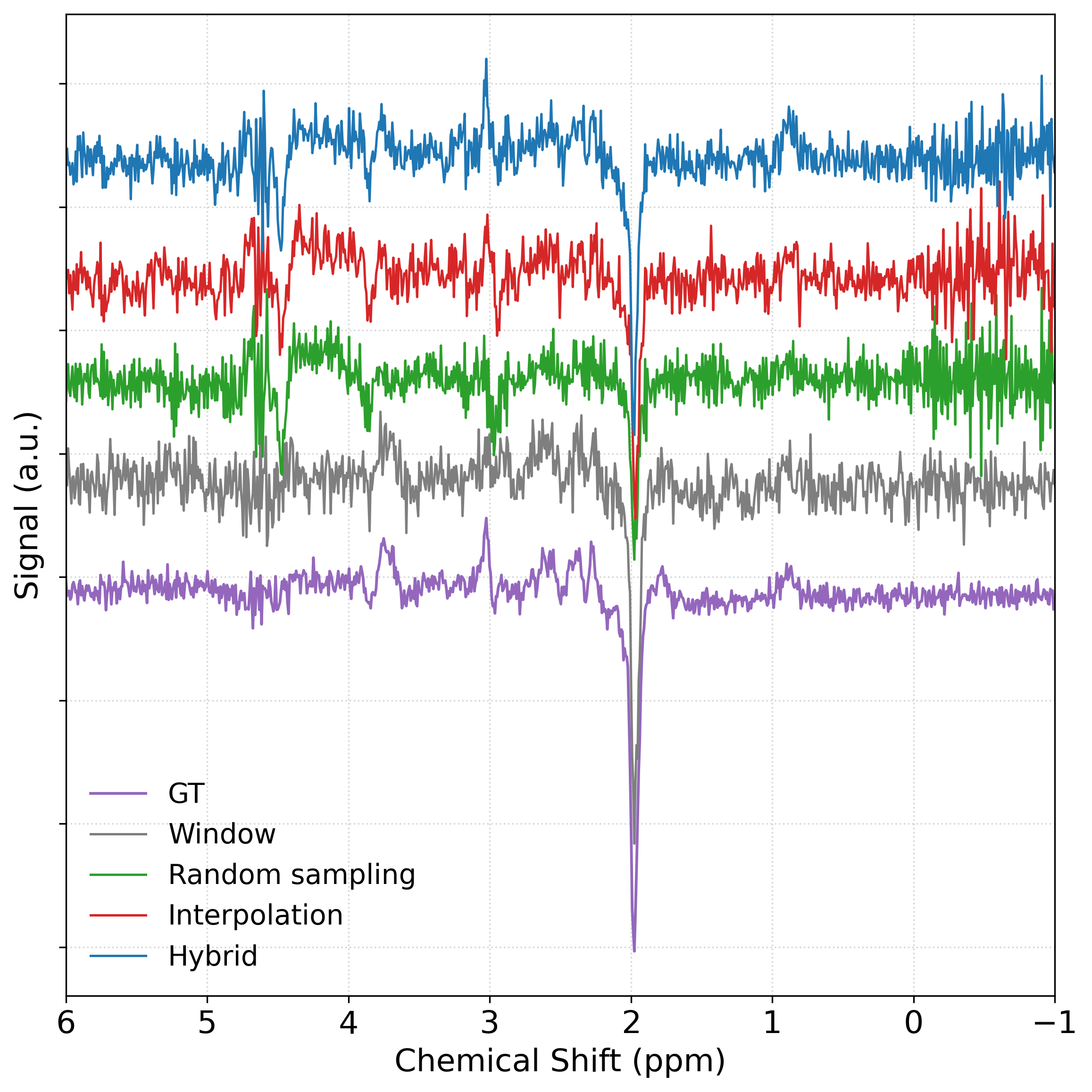}
        \caption{Difference (OFF-ON) spectra for all methods, subject C039.}
    \end{subfigure}

    \caption{Comparison of OFF, ON, and difference (OFF–ON) spectra across different data generation methods for two representative subjects. The top row (a-c) shows spectra from subject C008 (subset transient \#6), and the bottom row (d-f) shows spectra from subject C039 (subset transient \#6). For each subject, OFF spectra, ON spectra, and the corresponding difference spectra are shown from left to right.}
    \label{fig:generative_comparison}
\end{figure*}

Figure~\ref{fig:generative_comparison} compares the spectra generated by the different methods for two representative subjects and subsets of transients. In the top row (a–c), generating additional transients using the \ac{vae} results in an increased \ac{snr} compared to the original window-based spectra. Consequently, the \ac{gaba} and \ac{glx} peaks in the difference spectrum (Figure ~\ref{fig:generative_comparison}c) are more pronounced and more comparable to the \ac{gt} spectrum for all three generative methods.

In contrast, the bottom row (d–f) shows a case in which the spectra generated by the \ac{vae}-based methods deviate from the \ac{gt}. In this example, the OFF, ON, and difference spectra generated by all three methods exhibit increased noise and reduced agreement with the \ac{gt} compared to the example shown in the top row. Moreover, the synthetic spectra exhibit an \ac{naa} signal ($\sim2$ ppm) in the ON acquisition (Figure ~\ref{fig:generative_comparison}e), which is not expected for this experiment.

\begin{figure*}
    \centering
    \includegraphics[width=\linewidth]{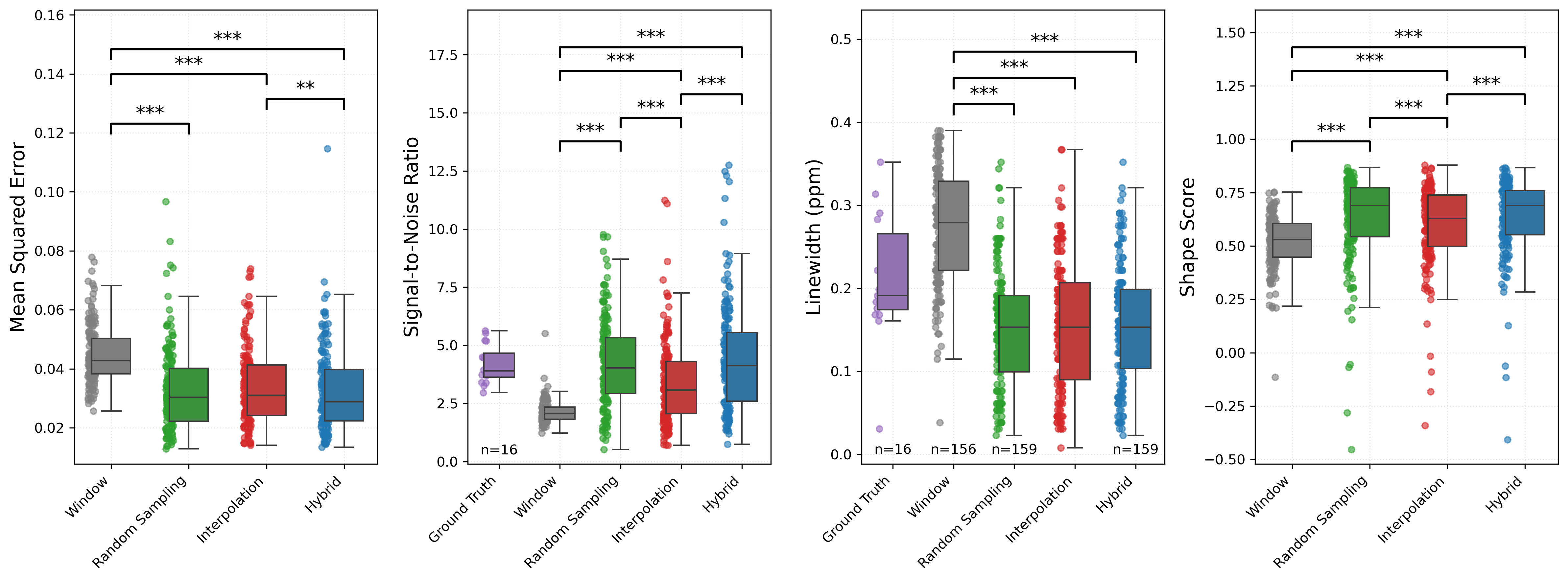}
    \caption{Overview of quantitative metrics for the selected windows (gray) and the different data generation methods: random sampling (green), interpolation (red), and hybrid generation (blue). Metrics are computed from the difference spectra. For the \ac{snr} and linewidth metrics, \ac{gt} values are shown as reference. Statistically significant differences are indicated (*** $p<0.001$, ** $p<0.01$, * $p<0.05$). Boxplots with fewer than the nominal $16 \times 10 = 160$ datapoints are explicitly marked.}
    \label{fig:metric_comparison}
\end{figure*}

Figure~\ref{fig:metric_comparison} presents a quantitative comparison of the metrics introduced in Section~\ref{subsubsec:application_based_evaluation}, calculated from the difference spectra. For the \ac{snr} and linewidth metrics, the corresponding \ac{gt} values are included as reference. In general, each boxplot comprises $16$ subjects across $10$ subsets of transients, resulting in $160$ datapoints, except for the \ac{gt}. In some cases, the linewidth could not be reliably calculated, leading to a reduced number of datapoints for this metric

Across all metrics, statistically significant differences are observed between the window-based spectra and all three data generation methods. In addition, significant differences are observed between the generative methods themselves, with a lower \ac{mse} for the hybrid method compared to interpolation, and lower \ac{snr} and shape scores for the interpolation method compared to both random sampling and hybrid generation.

\begin{figure*}
    \centering
    
    \begin{subfigure}[t]{\linewidth}
        \centering
        \includegraphics[width=\linewidth]{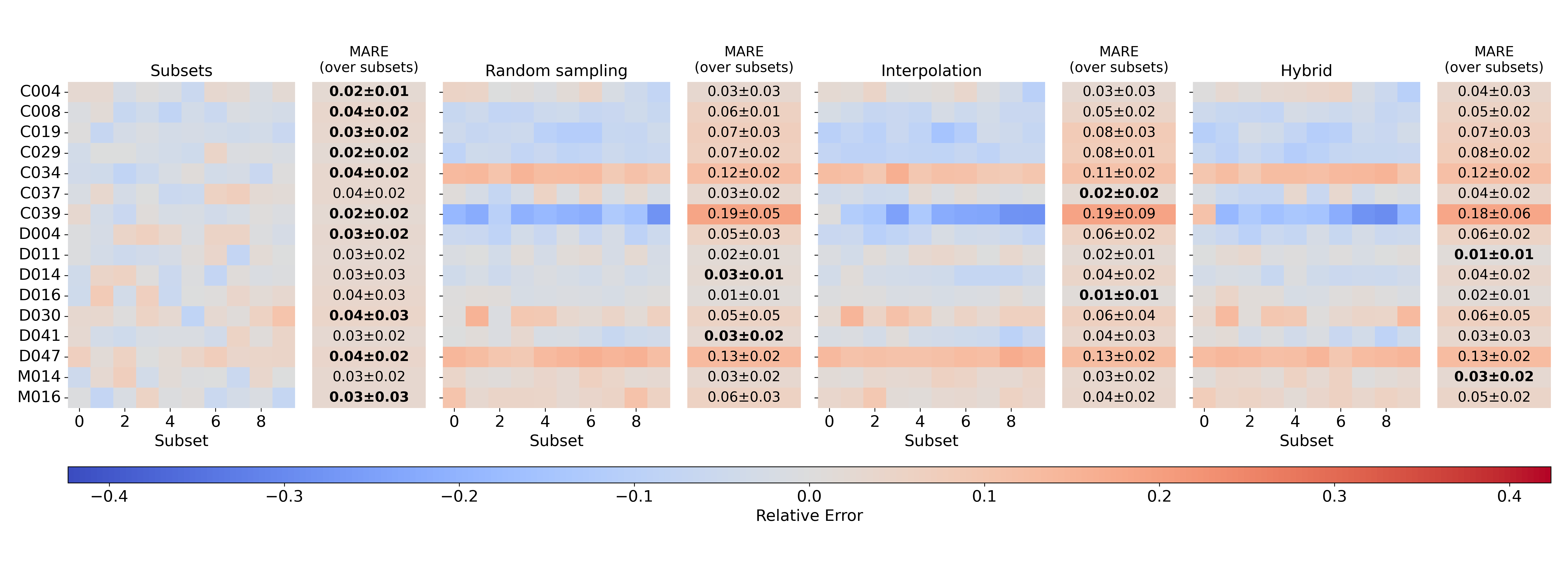}
        \caption{\ac{tnaa} quantification results for the OFF spectra.}
        \label{fig:quantification_results_off_tnaa}
    \end{subfigure}
    
    
    \begin{subfigure}[t]{\linewidth}
        \centering
        \includegraphics[width=\linewidth]{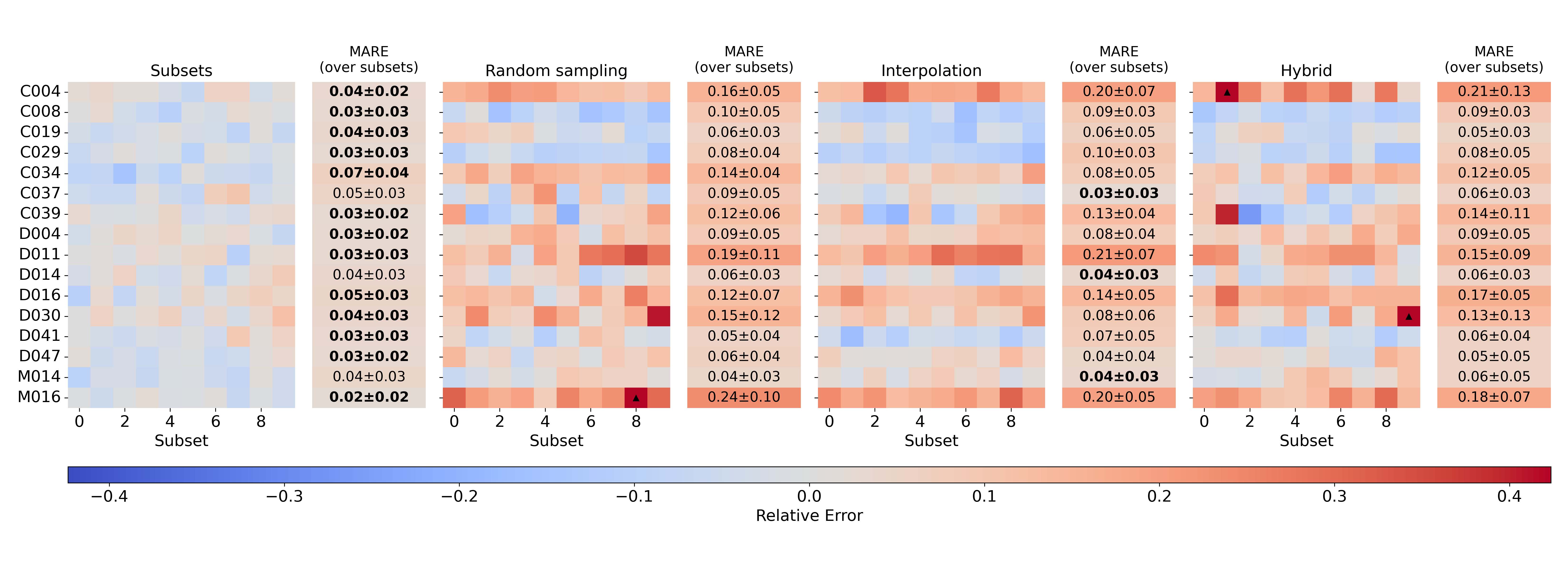}
        \caption{\ac{tcr} quantification results for the OFF spectra.}
        \label{fig:quantification_results_off_tcr}
    \end{subfigure}
    
    \caption{Heatmaps showing the relative concentration error with respect to the \ac{gt} test spectra for the OFF spectra. For each subject, the relative error is shown per window for the different reconstruction methods. The adjacent column summarizes the \ac{mare} across all windows for each subject. Values printed in bold indicate the method yielding the lowest \ac{mare} for that subject. The color scale is shared across methods. The $\blacktriangle$ symbols indicate values that exceed the displayed range.}
    \label{fig:quantification_results_off}
\end{figure*}

\begin{figure*}
    \centering
    
    \begin{subfigure}[t]{\linewidth}
        \centering
        \includegraphics[width=\linewidth]{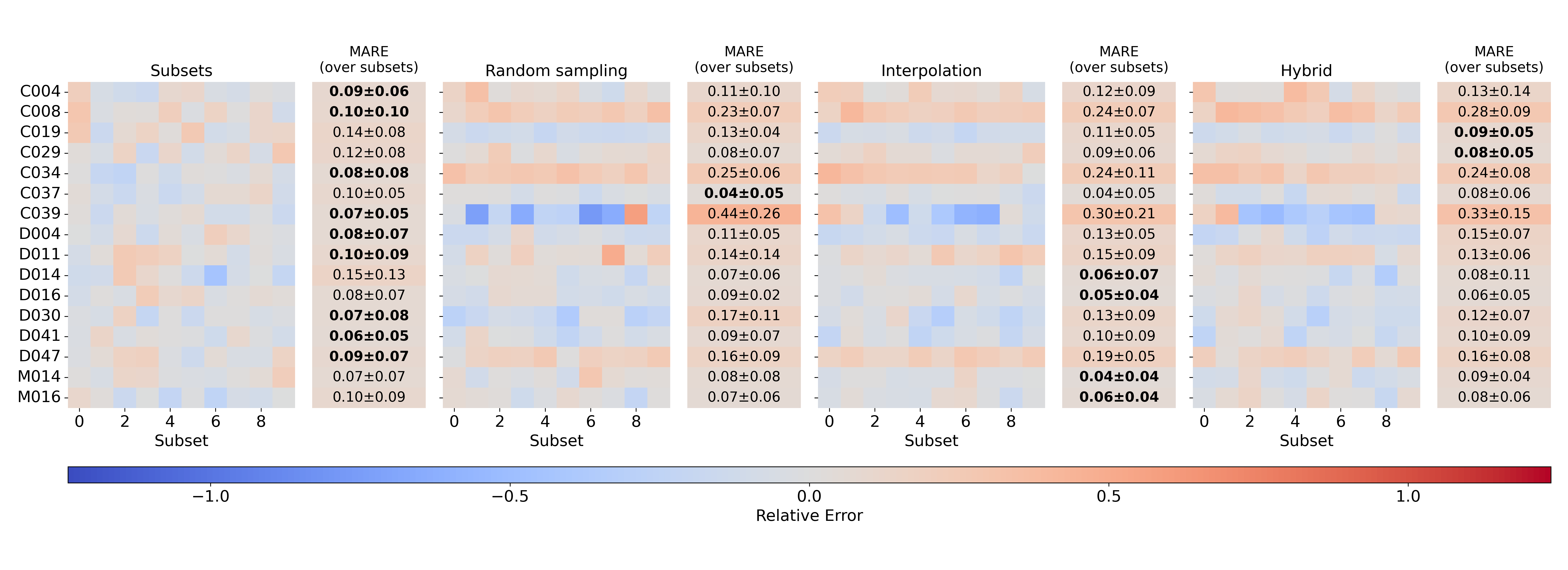}
        \caption{\ac{glx} quantification results for the difference spectra.}
        \label{fig:quantification_results_diff_glx}
    \end{subfigure}
    
    
    \begin{subfigure}[t]{\linewidth}
        \centering
        \includegraphics[width=\linewidth]{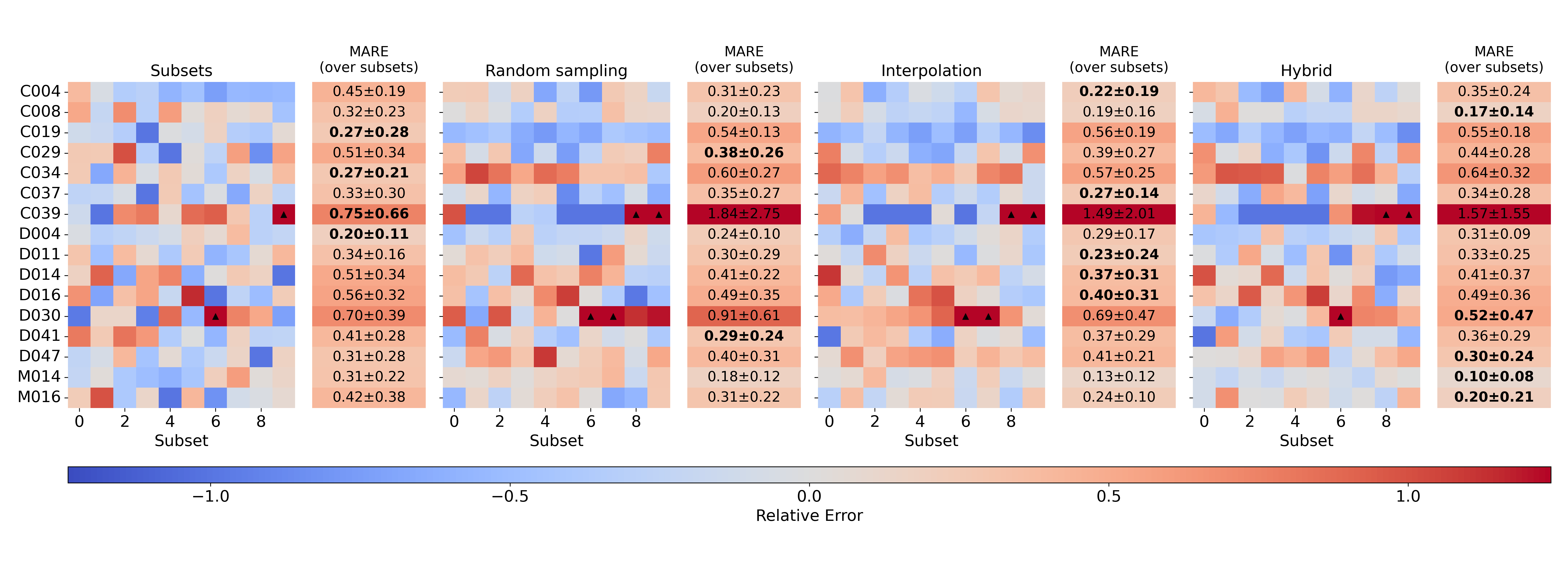}
        \caption{\ac{gaba} quantification results for the difference spectra.}
        \label{fig:quantification_results_diff_gaba}
    \end{subfigure}
    
    \caption{Heatmaps showing the relative concentration error with respect to the \ac{gt} test spectra for the difference spectra. For each subject, the relative error is shown per subset of transients for the different reconstruction methods. The adjacent column summarizes the \ac{mare} across all subsets for each subject. Values printed in bold indicate the method yielding the lowest \ac{mare} for that subject. The color scale is shared across methods. The $\blacktriangle$ symbols indicate values that exceed the displayed range.}
    \label{fig:quantification_results_diff}
\end{figure*}

In Figures \ref{fig:quantification_results_off} and \ref{fig:quantification_results_diff}, the quantification results for the OFF and difference spectra are shown as heatmaps, indicating the relative concentration error with respect to the \ac{gt} test spectra. For each subject, the relative error is reported per subset of transients, and the corresponding \ac{mare} across all subsets is summarized in an adjacent column. The \ac{mare} values shown in bold indicate the method yielding the lowest average deviation from the \ac{gt} quantification for that subject.

Figure \ref{fig:quantification_results_off} shows the quantification results for the \ac{tnaa} and \ac{tcr} metabolites derived from the OFF spectra. For \ac{tnaa}, the relative errors are generally small across subjects and subsets of transients, with most values remaining close to zero. Nevertheless, for all three generative methods, several subjects and subsets exhibit larger deviations. In particular, subject C039 shows repeated underestimations of the \ac{gt} amplitudes across subsets, while subjects C034 and D047 display more pronounced overestimations. When summarizing the errors across subsets, the lowest \ac{mare} values are most frequently observed for the original subset-based spectra. A similar pattern is observed for \ac{tcr}, for which the subset-based spectra again yield the lowest \ac{mare} values, whereas the generative methods tend to show systematic overestimations and increased variability.

The quantification results for the difference spectra are shown in Figure \ref{fig:quantification_results_diff} for the \ac{glx} and \ac{gaba} metabolites. For \ac{glx}, the original subsets again result in the lowest \ac{mare} values for most subjects, although for specific subjects and subsets the generative methods achieve lower relative errors. The relative errors for \ac{gaba} are larger compared to the other metabolites. Even for the original subset-based spectra, substantial deviations are present, with relative errors exceeding 1.0 for several subsets. While the \ac{mare} values remain high overall, the generative methods yield quantification results that are closer to the \ac{gt} for a majority of subjects, with the interpolation and hybrid approaches each producing the lowest \ac{mare} values for five subjects, and the random sampling approach for two subjects. For both \ac{glx} and \ac{gaba}, subject C039 again exhibits the largest deviations for the generative methods, consistent with the observations for the OFF spectra.

\section{Discussion}\label{sec:discussion}
The developed \ac{vae} model accurately reconstructs the dominant spectral patterns present in MRS data, but it has difficulty capturing stochastic components. In particular, the model does not accurately reproduce noise, which is consistent with its learning objective. During training, the \ac{vae} is optimized to encode and reconstruct systematic and recurrent spectral features, which are represented in the latent space. Random noise, by definition, lacks consistent structure across the spectral range and therefore cannot be effectively encoded. As a result, the decoder primarily reconstructs the underlying signal while largely suppressing stochastic fluctuations. This behavior is reflected in the comparable linewidths but differing \ac{snr} levels between the in-vivo and reconstructed spectra, as shown in Section~\ref{subsec:res_generative_performance}. Importantly, this limitation does not reduce the practical usefulness of the generated data if lower \ac{snr} is required, since noise can be added post-hoc to achieve desired \ac{snr} levels in a controlled manner.

A similar reconstruction challenge is observed for the residual water signal. Although this signal is not inherently stochastic, its amplitude, phase, and lineshape vary substantially across the dataset due to differences in water suppression efficiency and local field conditions, and its presence is not consistent across all spectra. In contrast to metabolite resonances, which are constrained by fixed chemical shifts and coupling patterns, residual water exhibits a much larger dynamic range and less structured variability, making it more difficult for the model to reproduce consistently. Consequently, the residual water component is not reliably reconstructed in all cases, as illustrated in Figure~\ref{fig:reconstructions}. Despite these imperfections, the synthetic spectra preserve the essential characteristics of the in-vivo data. As shown in Figure~\ref{fig:umap}, the generated spectra occupy the same low-dimensional feature space as the in-vivo spectra, indicating that the \ac{vae} captures the global metabolite-driven spectral structure of the in-vivo data, even though stochastic noise characteristics and artifact-related variability (e.g., baseline distortions or residual water) are not fully reproduced.

The \ac{umap} visualizations in Figure~\ref{fig:umap} further reveal clustering patterns corresponding to individual subjects in the dataset. While most subjects form clearly separated clusters in both the in-vivo and synthetic data, some datapoints lie in close proximity and partially overlap, indicating high feature similarity. This observation is particularly relevant for the random latent-space sampling strategy used to generate synthetic spectra. Assuming that spectra with similar structure are located near each other in latent space, excessive perturbations during sampling may cause transitions across cluster boundaries. As a result, features from different conditions—such as ON and OFF spectra, or even spectra from different subjects—may become inadvertently mixed. An example of this effect is shown in Figure~\ref{fig:generative_comparison}e, where synthetic ON spectra exhibit a detectable \ac{naa} signal that is not expected to be present. Although this behavior is not observed consistently across all subjects, it highlights the importance of carefully controlling latent-space sampling and implementing appropriate quality control when synthetic spectra are used in downstream processing pipelines.

A strength of the proposed approach is its ability to model heterogeneous data within a single generative framework. The \ac{vae} is trained jointly on spectra from different subject groups (diabetes, metabolic syndrome, and healthy controls) as well as ON and OFF transients, yet was still able to capture meaningful spectral structure. This is reflected in both the qualitative reconstructions and the overlap between in-vivo and synthetic spectra in the \ac{umap} feature space, as well as in the application-based metrics. These findings suggest that the model does not collapse distinct spectral characteristics despite the increased variability in the training data, and is capable of learning shared representations across conditions.

The application-based evaluation demonstrates improved performance across all assessed signal quality metrics, as shown in Figure~\ref{fig:metric_comparison}. The inclusion of synthetic spectra leads to improvements in \ac{mse}, \ac{snr}, linewidth, and shape score for all three data generation methods when compared to the subset-only approach. These results indicate that augmenting a limited subset of transients with synthetic data can enhance overall spectral quality. However, differences between synthetic and \ac{gt} spectra are observed for the \ac{snr} and linewidth metrics. The discrepancy in \ac{snr} is expected, given the previously discussed tendency of the \ac{vae} to omit stochastic noise. The observation of reduced linewidth in the synthetic spectra is more notable and likely reflects the fact that the generation process is conditioned on only two transients, preventing the model from capturing frequency variations present across the full acquisition. While this results in visually cleaner spectra, it may bias downstream applications that are sensitive to linewidth or frequency variability.

The final part of the evaluation examines the quantification performance of the OFF and difference spectra derived from the \ac{gt} and synthetic data. For the OFF spectra, the quantification results for \ac{tnaa} and \ac{tcr} (Figure~\ref{fig:quantification_results_off}) show clear differences between the subset-only spectra and the three synthetic data generation methods. While the subset-only approach yields low relative errors (\ac{mare} $\leq$ 0.07, and $\leq$ 0.04 in most cases), augmenting these subsets with synthetic spectra substantially increases errors, reaching up to 0.24 in the most extreme case. This effect is particularly pronounced for \ac{tcr}, indicating that augmenting a small subset of transients with synthetic spectra can adversely affect metabolite quantification, despite improvements in signal quality metrics. Although uncertainties in the quantification algorithm and in the definition of the \ac{gt} concentrations cannot be fully excluded, these results suggest that the \ac{vae} does not fully preserve the spectral features required for accurate concentration estimation.

A similar pattern is observed for the difference spectra (Figure~\ref{fig:quantification_results_diff}). The \ac{glx} amplitudes obtained from the subset-only spectra are generally closer to the \ac{gt} values than those derived from synthetic spectra. In contrast, the quantification of \ac{gaba} remains challenging across all approaches, including the subset-only spectra, reflecting the inherent difficulty of reliable \ac{gaba} estimation. Taken together, these results highlight that while synthetic data generation can improve spectral appearance and stability, it does not necessarily preserve quantitative metabolite accuracy, particularly for applications that rely on absolute concentration estimates.

Importantly, the synthetic spectra primarily expand interspectral variability and not intersubject variability. With a limited number of subjects used to train the \ac{vae}, additional generated spectra do not introduce new biological differences between individuals. Rather, they provide model-based samples that remain constrained to the local spectral manifold of the observed subjects. Consequently, synthetic spectra can enrich spectral level variation within subjects, but they should not be interpreted as increasing the effective biological diversity of the cohort. This distinction is particularly important for analyses that depend on realistic intersubject variation.

Beyond the specific performance characteristics of the proposed \ac{vae}, an important contribution of this work lies in the multi-faceted nature of the evaluation strategy applied to generative modeling in MRS. Generative approaches in spectroscopy are often assessed primarily through visual inspection or a limited set of signal quality metrics, which may not fully capture their suitability for downstream applications. In contrast, this study combines spectral reconstruction metrics, low-dimensional feature space analysis, application-based evaluation, and metabolite quantification to assess both the strengths and limitations of the generated data. By explicitly identifying scenarios in which synthetic spectra improve signal characteristics as well as cases where they introduce subtle but consequential deviations, this work underscores the necessity of comprehensive, application-aware validation of generative models in spectroscopy.

While the proposed method may not be optimal for applications that require accurate absolute metabolite quantification, the ability to generate realistic spectra conditioned on limited input data may be advantageous for other downstream tasks. In particular, generative augmentation could be used to balance or oversample specific classes in classification settings, where relative spectral differences and shared structure are more important than precise concentration estimates. Such applications could benefit from the increased data diversity provided by synthetic spectra without requiring strict preservation of absolute metabolite levels. However, the dataset used in this study is not well suited for evaluating this scenario, as differences between subject groups are subtle at the level of individual spectra and primarily become apparent only when aggregating results at the group level \cite{vanbusselIncreasedGABAConcentrations2016}. Exploring the potential of the proposed approach for classification or other discriminative tasks therefore remains an important direction for future work.

\section{Conclusion}\label{sec:conclusion}
In this study, we developed a \ac{vae}-based generative model capable of synthesizing realistic proton brain MRS spectra, capturing the dominant spectral patterns of an in-vivo dataset in low-dimensional features The model preserves global spectral characteristics, as confirmed by reconstruction performance, \ac{umap} embeddings, and quantitative metrics, and it improves signal quality metrics across multiple synthetic data generation approaches. However, limitations remain, including inconsistent reconstruction of stochastic components such as noise and residual water signals, as well as reduced accuracy in metabolite quantification. Beyond these specific results, this work demonstrates a comprehensive evaluation framework for generative MRS models, highlighting both their potential to augment datasets and the importance of careful validation for quantitative or downstream analyses. Future studies could explore these applications further, and extend the model to larger datasets or alternative acquisition schemes, building on this framework. Together, these results provide a foundation for the responsible and effective use of generative \ac{ai} in MRS research.

\bibliographystyle{unsrt}  
\bibliography{refs} 

@article{linMinimumReportingStandards2021,
  title = {Minimum {{Reporting Standards}} for in Vivo {{Magnetic Resonance Spectroscopy}} ({{MRSinMRS}}): {{Experts}}' Consensus Recommendations},
  shorttitle = {Minimum {{Reporting Standards}} for in Vivo {{Magnetic Resonance Spectroscopy}} ({{MRSinMRS}})},
  author = {Lin, Alexander and Andronesi, Ovidiu and Bogner, Wolfgang and Choi, In-Young and Coello, Eduardo and Cudalbu, Cristina and Juchem, Christoph and Kemp, Graham J. and Kreis, Roland and Kr{\v s}{\v s}{\'a}k, Martin and Lee, Phil and Maudsley, Andrew A. and Meyerspeer, Martin and Mlynarik, Vladamir and Near, Jamie and {\"O}z, G{\"u}lin and Peek, Aimie L. and Puts, Nicolaas A. and Ratai, Eva-Maria and Tk{\'a}{\v c}, Ivan and Mullins, Paul G. and Spectroscopy, Experts' Working Group on Reporting Standards for MR},
  year = 2021,
  journal = {NMR in Biomedicine},
  volume = {34},
  number = {5},
  pages = {e4484},
  issn = {1099-1492},
  doi = {10.1002/nbm.4484},
  urldate = {2022-02-24},
  langid = {english}
}

@article{vandesandeDigitalPhantomMR2026,
  title = {A {{Digital Phantom}} for {{MR Spectroscopy Data Simulation}}},
  author = {{van de Sande}, D. M. J. and Gudmundson, A. T. and {Murali-Manohar}, S. and {Davies-Jenkins}, C. W. and Simicic, D. and Simegn, G. and {\"O}zdemir, {\.I}. and Amirrajab, S. and Merkofer, J. P. and Z{\"o}llner, H. J. and Oeltzschner, G. and Edden, R. a. E.},
  year = 2026,
  journal = {Magnetic Resonance in Medicine},
  volume = {95},
  number = {3},
  pages = {1345--1359},
  issn = {1522-2594},
  doi = {10.1002/mrm.70138},
  urldate = {2026-01-20},
  copyright = {\copyright{} 2025 The Author(s). Magnetic Resonance in Medicine published by Wiley Periodicals LLC on behalf of International Society for Magnetic Resonance in Medicine.},
  langid = {english}
}

@article{vandesandeReviewMachineLearning2023,
  title = {A Review of Machine Learning Applications for the Proton {{MR}} Spectroscopy Workflow},
  author = {{van de Sande}, Dennis M. J. and Merkofer, Julian P. and Amirrajab, Sina and Veta, Mitko and {van Sloun}, Ruud J. G. and Versluis, Maarten J. and Jansen, Jacobus F. A. and {van den Brink}, Johan S. and Breeuwer, Marcel},
  year = 2023,
  journal = {Magnetic Resonance in Medicine},
  volume = {90},
  number = {4},
  pages = {1253--1270},
  issn = {1522-2594},
  doi = {10.1002/mrm.29793},
  urldate = {2023-11-07},
  copyright = {\copyright{} 2023 The Authors. Magnetic Resonance in Medicine published by Wiley Periodicals LLC on behalf of International Society for Magnetic Resonance in Medicine.},
  langid = {english}
}

@article{bertoResults2023ISBI2024,
  title = {Results of the 2023 {{ISBI}} Challenge to Reduce {{GABA-edited MRS}} Acquisition Time},
  author = {Berto, Rodrigo Pommot and Bugler, Hanna and Dias, Gabriel and Oliveira, Mateus and Ueda, Lucas and Dertkigil, Sergio and Costa, Paula D. P. and Rittner, Leticia and Merkofer, Julian P. and {van de Sande}, Dennis M. J. and Amirrajab, Sina and Drenthen, Gerhard S. and Veta, Mitko and Jansen, Jacobus F. A. and Breeuwer, Marcel and {van Sloun}, Ruud J. G. and Qayyum, Abdul and Rodero, Cristobal and Niederer, Steven and Souza, Roberto and Harris, Ashley D.},
  year = 2024,
  month = apr,
  journal = {Magnetic Resonance Materials in Physics, Biology and Medicine},
  issn = {1352-8661},
  doi = {10.1007/s10334-024-01156-9},
  urldate = {2024-05-21},
  langid = {english}
}

@article{mulkernDensityMatrixCalculations1994,
  title = {Density Matrix Calculations of {{AB}} Spectra from Multipulse Sequences: {{Quantum}} Mechanics {{meetsIn}} Vivo Spectroscopy},
  shorttitle = {Density Matrix Calculations of {{AB}} Spectra from Multipulse Sequences},
  author = {Mulkern, Robert and Bowers, John},
  year = 1994,
  month = jan,
  journal = {Concepts in Magnetic Resonance},
  volume = {6},
  number = {1},
  pages = {1--23},
  issn = {10437347, 10990534},
  doi = {10.1002/cmr.1820060102},
  urldate = {2022-02-11},
  langid = {english}
}

@article{clarkeNIfTIMRSStandardData2022,
  title = {{{NIfTI-MRS}}: {{A}} Standard Data Format for Magnetic Resonance Spectroscopy},
  shorttitle = {{{NIfTI-MRS}}},
  author = {Clarke, William T. and Bell, Tiffany K. and Emir, Uzay E. and Mikkelsen, Mark and Oeltzschner, Georg and Shamaei, Amirmohammad and Soher, Brian J. and Wilson, Martin},
  year = 2022,
  journal = {Magnetic Resonance in Medicine},
  volume = {88},
  number = {6},
  pages = {2358--2370},
  issn = {1522-2594},
  doi = {10.1002/mrm.29418},
  urldate = {2024-07-22},
  copyright = {\copyright{} 2022 The Authors. Magnetic Resonance in Medicine published by Wiley Periodicals LLC on behalf of International Society for Magnetic Resonance in Medicine.},
  langid = {english}
}

@article{clarkeFSLMRSEndtoendSpectroscopy2021,
  title = {{{FSL-MRS}}: {{An}} End-to-End Spectroscopy Analysis Package},
  shorttitle = {{{FSL-MRS}}},
  author = {Clarke, William T. and Stagg, Charlotte J. and Jbabdi, Saad},
  year = 2021,
  journal = {Magnetic Resonance in Medicine},
  volume = {85},
  number = {6},
  pages = {2950--2964},
  issn = {1522-2594},
  doi = {10.1002/mrm.28630},
  urldate = {2022-03-15},
  langid = {english}
}

@article{chenReviewProspectDeep2020,
    author = {Chen, Dicheng and Wang, Zi and Guo, Di and Orekhov, Vladislav and Qu, Xiaobo},
    title = {Review and Prospect: Deep Learning in Nuclear Magnetic Resonance Spectroscopy},
    journal = {Chemistry – A European Journal},
    volume = {26},
    number = {46},
    pages = {10391-10401},
    doi = {https://doi.org/10.1002/chem.202000246},
    url = {https://chemistry-europe.onlinelibrary.wiley.com/doi/abs/10.1002/chem.202000246},
    eprint = {https://chemistry-europe.onlinelibrary.wiley.com/doi/pdf/10.1002/chem.202000246},
    year = {2020}
}

@article{simpsonAdvancedProcessingSimulation2017,
  title = {Advanced Processing and Simulation of {{MRS}} Data Using the {{FID}} Appliance ({{FID-A}})---{{An}} Open Source, {{MATLAB-based}} Toolkit},
  author = {Simpson, Robin and Devenyi, Gabriel A. and Jezzard, Peter and Hennessy, T. Jay and Near, Jamie},
  year = 2017,
  journal = {Magnetic Resonance in Medicine},
  volume = {77},
  number = {1},
  pages = {23--33},
  issn = {1522-2594},
  doi = {10.1002/mrm.26091},
  urldate = {2022-02-10},
  langid = {english}
}

@article{soherVespaIntegratedApplications2023,
  title = {Vespa: {{Integrated}} Applications for {{RF}} Pulse Design, Spectral Simulation and {{MRS}} Data Analysis},
  shorttitle = {Vespa},
  author = {Soher, Brian J. and Semanchuk, Philip and Todd, David and Ji, Xiao and Deelchand, Dinesh and Joers, James and Oz, Gulin and Young, Karl},
  year = 2023,
  journal = {Magnetic Resonance in Medicine},
  volume = {90},
  number = {3},
  pages = {823--838},
  issn = {1522-2594},
  doi = {10.1002/mrm.29686},
  urldate = {2025-08-14},
  copyright = {\copyright{} 2023 International Society for Magnetic Resonance in Medicine},
  langid = {english}
}

@article{lamasterMRSSimOpenSourceFramework2025,
  title = {{{MRS-Sim}}: {{Open-Source Framework}} for {{Simulating In Vivo-Like Magnetic Resonance Spectra}}},
  shorttitle = {{{MRS-Sim}}},
  author = {LaMaster, John and Oeltzschner, Georg and Li, Yan},
  year = 2025,
  journal = {NMR in Biomedicine},
  volume = {38},
  number = {10},
  pages = {e70130},
  issn = {1099-1492},
  doi = {10.1002/nbm.70130},
  urldate = {2026-01-20},
  copyright = {\copyright{} 2025 The Author(s). NMR in Biomedicine published by John Wiley \& Sons Ltd.},
  langid = {english}
}

@article{flanaganExploringGenerativeArtificial2025,
  title = {Exploring {{Generative Artificial Intelligence}} and {{Data Augmentation Techniques}} for {{Spectroscopy Analysis}}},
  author = {Flanagan, Aaron R. and Dalal, Dhairya and Glavin, Frank G.},
  year = 2025,
  month = jul,
  journal = {Chemical Reviews},
  volume = {125},
  number = {13},
  pages = {6130--6155},
  publisher = {American Chemical Society},
  issn = {0009-2665},
  doi = {10.1021/acs.chemrev.4c00815},
  urldate = {2025-11-17}
}

@article{jangUnsupervisedAnomalyDetection2021,
  title = {Unsupervised Anomaly Detection Using Generative Adversarial Networks in {{1H-MRS}} of the Brain},
  author = {Jang, Joon and Lee, Hyeong Hun and Park, Ji-Ae and Kim, Hyeonjin},
  year = 2021,
  month = apr,
  journal = {Journal of Magnetic Resonance},
  volume = {325},
  pages = {106936},
  issn = {1090-7807},
  doi = {10.1016/j.jmr.2021.106936},
  urldate = {2026-01-20}
}

@article{kebailiDeepLearningApproaches2023,
  title = {Deep {{Learning Approaches}} for {{Data Augmentation}} in {{Medical Imaging}}: {{A Review}}},
  shorttitle = {Deep {{Learning Approaches}} for {{Data Augmentation}} in {{Medical Imaging}}},
  author = {Kebaili, Aghiles and {Lapuyade-Lahorgue}, J{\'e}r{\^o}me and Ruan, Su},
  year = 2023,
  month = apr,
  journal = {Journal of Imaging},
  volume = {9},
  number = {4},
  pages = {81},
  issn = {2313-433X},
  doi = {10.3390/jimaging9040081},
  langid = {english},
  pmcid = {PMC10144738},
  pmid = {37103232}
}

@misc{kingmaAutoEncodingVariationalBayes2014,
  author       = {Kingma, Diederik P. and Welling, Max},
  title        = {Auto-Encoding Variational Bayes},
  year         = {2014},
  month        = may,
  howpublished = {\url{https://arxiv.org/abs/1312.6114}},
  note         = {Preprint, arXiv:1312.6114},
}

@article{maruyamaGeneratingSyntheticMR2025,
  title = {Generating {{Synthetic MR Spectroscopic Imaging Data}} with {{Generative Adversarial Networks}} to {{Train Machine Learning Models}}},
  author = {Maruyama, Shuki and Takeshima, Hidenori},
  year = 2025,
  journal = {Magnetic Resonance in Medical Sciences},
  volume = {24},
  number = {4},
  pages = {n/a},
  issn = {1347-3182, 1880-2206},
  doi = {10.2463/mrms.mp.2023-0125},
  urldate = {2025-10-13},
  langid = {english}
}

@misc{mcinnesUMAPUniformManifold2020,
  title = {{{UMAP}}: {{Uniform Manifold Approximation}} and {{Projection}} for {{Dimension Reduction}}},
  shorttitle = {{{UMAP}}},
  author = {McInnes, Leland and Healy, John and Melville, James},
  date = {2020-09-18},
  eprint = {1802.03426},
  eprinttype = {arXiv},
  eprintclass = {stat},
  doi = {10.48550/arXiv.1802.03426},
  url = {http://arxiv.org/abs/1802.03426},
  urldate = {2025-11-04},
  pubstate = {prepublished},
  howpublished = {\url{http://arxiv.org/abs/1802.03426}},
}

@article{oeltzschnerOspreyOpensourceProcessing2020,
  title = {Osprey: {{Open-source}} Processing, Reconstruction \& Estimation of Magnetic Resonance Spectroscopy Data},
  shorttitle = {Osprey},
  author = {Oeltzschner, Georg and Z{\"o}llner, Helge J. and Hui, Steve C. N. and Mikkelsen, Mark and Saleh, Muhammad G. and Tapper, Sofie and Edden, Richard A. E.},
  year = 2020,
  month = sep,
  journal = {Journal of Neuroscience Methods},
  volume = {343},
  pages = {108827},
  issn = {0165-0270},
  doi = {10.1016/j.jneumeth.2020.108827},
  urldate = {2022-02-16},
  langid = {english}
}

@inproceedings{olliverreGeneratingMagneticResonance2018,
  author = {Olliverre, N. and Yang, G. and Slabaugh, G. and Reyes-Aldasoro, C. C. and Alonso, E.},
  title = {Generating Magnetic Resonance Spectroscopy Imaging Data of Brain Tumours from Linear, Non-linear and Deep Learning Models},
  booktitle = {Simulation and Synthesis in Medical Imaging},
  editor = {Gooya, A. and Goksel, O. and Oguz, I. and Burgos, N.},
  series = {Lecture Notes in Computer Science},
  volume = {11037},
  publisher = {Springer},
  location = {Cham},
  year = {2018},
  doi = {10.1007/978-3-030-00536-8_14},
  organization = {SASHIMI 2018 / MICCAI Society},
}

@misc{raisExploringVariationalAutoencoders2024,
  author       = {Rais, Khadija and Amroune, Mohamed and Benmachiche, Abdelmadjid and Haouam, Mohamed Yassine},
  title        = {Exploring Variational Autoencoders for Medical Image Generation: A Comprehensive Study},
  year         = {2024},
  month        = nov,
  howpublished = {\url{https://arxiv.org/abs/2411.07348}},
  note         = {Preprint, arXiv:2411.07348},
}

@article{schramMaastrichtStudyExtensive2014,
  title = {The {{Maastricht Study}}: An Extensive Phenotyping Study on Determinants of Type 2 Diabetes, Its Complications and Its Comorbidities},
  shorttitle = {The {{Maastricht Study}}},
  author = {Schram, Miranda T. and Sep, Simone J. S. and {van der Kallen}, Carla J. and Dagnelie, Pieter C. and Koster, Annemarie and Schaper, Nicolaas and Henry, Ronald M. A. and Stehouwer, Coen D. A.},
  year = 2014,
  month = jun,
  journal = {European Journal of Epidemiology},
  volume = {29},
  number = {6},
  pages = {439--451},
  issn = {1573-7284},
  doi = {10.1007/s10654-014-9889-0},
  urldate = {2024-02-21},
  langid = {english}
}

@article{vanbusselIncreasedGABAConcentrations2016,
  title = {Increased {{GABA}} Concentrations in Type 2 Diabetes Mellitus Are Related to Lower Cognitive Functioning},
  author = {{van Bussel}, Frank C. G. and Backes, Walter H. and Hofman, Paul A. M. and Puts, Nicolaas A. J. and Edden, Richard A. E. and {van Boxtel}, Martin P. J. and Schram, Miranda T. and Stehouwer, Coen D. A. and Wildberger, Joachim E. and Jansen, Jacobus F. A.},
  year = 2016,
  month = sep,
  journal = {Medicine},
  volume = {95},
  number = {36},
  pages = {e4803},
  issn = {1536-5964},
  doi = {10.1097/MD.0000000000004803},
  langid = {english},
  pmcid = {PMC5023915},
  pmid = {27603392}
}

\section*{Code Availability}
All code used in this study is publicly available in a GitHub repository. The repository link will be provided upon final publication.

\section*{Conflict of interest}
The authors declare no potential conflict of interests.

\appendix
\newpage
\section{MRS in MRS} \label{app:mrs_in_mrs}

This section provides transparency and reproducibility by adhering to the \ac{mrsinmrs} guidelines \cite{linMinimumReportingStandards2021}. The table details the hardware, acquisition, processing, and data quality parameters for this study.

\begin{table*}
    \caption{Minimum Reporting Standards for in-vivo Magnetic Resonance Spectroscopy (MRSinMRS) \cite{linMinimumReportingStandards2021} for used dataset. Details are retrieved from the data and the work from van Bussel et al. \cite{vanbusselIncreasedGABAConcentrations2016}} 
    \label{tab:mrsinmrs_data}
    \vspace{2mm}
    \begin{tabularx}{\textwidth}{@{} l >{\raggedright\arraybackslash}X @{}}
    \toprule
    \textbf{Header / SubHeader} & \textbf{Values} \\
    \midrule
    \textbf{Site (name or number)} &  \\
    \midrule
    \textbf{1. Hardware} & \\
    a. Field strength & 3T \\
    b. Manufacturer & Philips \\
    c. Model (software version if available) & Achieva TX \\
    d. RF coils: nuclei (transmit/receive), number of channels, type, body part & 1H, 32-element, head coil \\
    e. Additional hardware & - \\
    \midrule
    \textbf{2. Acquisition} & \\
    a. Pulse sequence & MEGA-PRESS\\
    b. Volume of interest (VOI) locations & occipital lobe \\
    c. Nominal VOI size &  30$\times$30$\times$30 mm$^3$\\
    d. Repetition time (TR), echo time (TE) & TR = 2000 ms, TE = 68 ms \\
    e. Total number of excitations or acquisitions per spectrum & 320, interleaved in 40 blocks \\
    f. Additional sequence parameters & 2000 Hz spectral width, 2048 spectral points, editing pulses at 1.9 (ON), and 7.46 ppm (OFF) \\
    g. Water suppression method & MOIST water suppression \\
    h. Shimming method, reference peak, and thresholds for “acceptance of shim” chosen & unknown \\
    i. Triggering or motion correction method & - \\
    \midrule
    \textbf{3. Data analysis methods and outputs} & \\
    a. Analysis software & In-house Python scripts \newline FSL-MRS\cite{clarkeFSLMRSEndtoendSpectroscopy2021} (2.1.20) \newline Osprey \cite{oeltzschnerOspreyOpensourceProcessing2020} (2.9.6)  \\
    b. Processing steps deviating from quoted reference or product & 
    Alignment of dynamics along time dimension (0.2–4.2 ppm); water referencing and eddy current correction (ECC) using reference data; frequency shift to 3.027 ppm reference peak; phase correction over 2.9–3.1 ppm.
     \\
    c. Output measure & Amplitudes/institutional units \\
    d. Quantification references and assumptions, fitting model assumptions & See Osprey job file in Appendix \ref{app:osprey_job} \\
    \midrule
    \textbf{4. Data quality} & \\
    a. Reported variables (SNR, linewidth (with reference peaks)) & SNR: $9.98 \pm 1.18$ (OFF) and $11.06 \pm 1.35$ (ON), FWHM: $0.057 \pm 0.017$ ppm (OFF) and $0.062 \pm 0.018$ ppm (ON), all referenced to \ac{cr}  \\
    b. Data exclusion criteria & - \\
    c. Quality measures of postprocessing model fitting & - \\
    d. Sample spectrum & See Figures \ref{fig:reconstructions} and \ref{fig:generative_comparison}\\
    \bottomrule
    \end{tabularx}
\end{table*}

\newpage

\section{Example MATLAB Osprey Job Script}\label{app:osprey_job}

\begin{figure*}[!h]

\lstset{
    language=Matlab,
    basicstyle=\ttfamily\small,
    keywordstyle=\color{blue}\bfseries,
    commentstyle=\color{gray},
    stringstyle=\color{purple},
    breaklines=true,
    frame=single,
    numbers=left,
    numberstyle=\tiny,
    stepnumber=1,
    tabsize=4,
    showstringspaces=false,
    captionpos=b
}

\begin{lstlisting}[caption={Osprey Job script used for the quantification of the in-vivo and synthetic spectra.}]

%% 1. SPECIFY SEQUENCE INFORMATION
seqType = 'MEGA';                     % Options: 'unedited', 'MEGA', 'HERMES', 'HERCULES'
editTarget = {'GABA'};                % Editing targets
dataScenario = 'invivo';              % Options: 'invivo', 'phantom', 'PRIAM', 'MRSI'

%% 2. DATA HANDLING AND MODELING OPTIONS
opts.SpecReg               = 'RobSpecReg';    % Spectral registration method
opts.SubSpecAlignment.mets = 'L2Norm';        % Sub-spectrum alignment
opts.ECC.raw               = 1;               % Eddy current correction
opts.ECC.mm                = 1;
opts.saveLCM               = 0;
opts.savejMRUI             = 0;
opts.saveVendor            = 0;
opts.saveNII               = 1;
opts.savePDF               = 0;
opts.exportParams.flag     = 0;
opts.exportParams.path     = '';               % Path to save parameters

opts.fit.method            = 'Osprey';        % Fitting algorithm
opts.fit.includeMetabs     = {'default'};      % Metabolites to include
opts.fit.style             = 'Separate';      % Fit style for edited datasets
opts.fit.range             = [0.5 4];         % Fitting range [ppm]
opts.fit.rangeWater        = [2.0 7.4];
opts.fit.bLineKnotSpace    = 0.55;
opts.fit.fitMM             = 1;               % Include macromolecules/lipids
opts.fit.coMM3             = '3to2MM';
opts.fit.FWHMcoMM3         = 14;
opts.fit.basisSetFile      = '/PATH/TO/BASISSET/basis_philips_megapress_gaba68.mat';

opts.img.deface            = 0;               % Optional: deface structural images

%% 3. SPECIFY MRS DATA AND STRUCTURAL IMAGING FILES
clear files files_ref files_w files_nii files_mm

root = '/PATH/TO/MRS_DATA';
model = 'MODEL_NAME';
subjects = {};
windows  = [];
types     = {'gt','window','interpolation','random_sampling','random_interpolation'};

files = getNiftiFiles(root, model, subjects, windows, types);

%% 4. SPECIFY STAT FILE
% Not used in this example

%% 5. SPECIFY OUTPUT FOLDER
outputFolder = fullfile('/PATH/TO/OUTPUT', model);
if ~exist(outputFolder, 'dir')
    mkdir(outputFolder);
end

% Save file list
fileListTxt = fullfile(outputFolder, 'file_list.txt');
fid = fopen(fileListTxt, 'w');
for i = 1:length(files)
    fprintf(fid, '%s\n', files{i});
end
fclose(fid);

\end{lstlisting}

\end{figure*}

\end{document}